\documentclass{aa}
\usepackage{graphicx}
\usepackage{txfonts}
\usepackage{natbib}
\bibpunct{(}{)}{;}{a}{}{,}
\usepackage{longtable}
\usepackage{lscape}
\usepackage{times}
\usepackage{textcomp}
\usepackage{mathrsfs}
\usepackage{rotating}

\begin{document}

\title{Magnetic fields of the W4 superbubble}
\subtitle{}
\author{X. Y.~Gao\inst{1,2}, W.~Reich\inst{2}, P.~Reich\inst{2},
  J. L.~Han\inst{1}, and R.~Kothes\inst{3}}
\titlerunning{W4 superbubble}
\authorrunning{Gao et al.}

\offprints{bearwards@gmail.com}

\institute{National Astronomical Observatories, Chinese Academy of
  Sciences, Jia-20 Datun Road, Chaoyang District, Beijing 100012, PR
  China \and Max-Planck-Institut f\"{u}r Radioastronomie, Auf dem
  H\"{u}gel 69, 53121 Bonn, Germany \and National Research Council of
  Canada, Dominion Radio Astrophysical Observatory, P.O. Box 248,
  Penticton BC, V2A 6J9, Canada}

\date{Received; accepted}

\abstract
{Superbubbles and supershells are the channels for transferring mass
  and energy from the Galactic disk to the halo. Magnetic fields are
  believed to play a vital role in their evolution.}
{We study the radio continuum and polarized emission properties of the
  W4 superbubble to determine its magnetic field strength.}
{New sensitive radio continuum observations were made at
  $\lambda$6\ cm, $\lambda$11\ cm, and $\lambda$21\ cm. The total
  intensity measurements were used to derive the radio spectrum of the
  W4 superbubble. The linear polarization data were analysed to
  determine the magnetic field properties within the bubble shells.}
{The observations show a multi-shell structure of the W4
  superbubble. A flat radio continuum spectrum that stems from
  optically thin thermal emission is derived from 1.4~GHz to 4.8~GHz.
  By fitting a passive Faraday screen model and considering the
  filling factor $f_{n_e}$, we obtain the thermal electron density
  n$_{e}$ = $1.0/\sqrt{f_{n_e}}$ ($\pm$5\%)~cm$^{-3}$ and the strength
  of the line-of-sight component of the magnetic field $B_{//}$ =
  $-5.0/\sqrt{f_{n_e}}$ ($\pm$10\%)~$\mu$G (i.e. pointing away from
  us) within the western shell of the W4 superbubble.  When the known
  tilted geometry of the W4 superbubble is considered, the total
  magnetic field B$_{tot}$ in its western shell is greater than
  12~$\mu$G. The electron density and the magnetic field are lower and
  weaker in the high-latitude parts of the superbubble. The rotation
  measure is found to be positive in the eastern shell but negative in
  the western shell of the W4 superbubble, which is consistent with
  the case that the magnetic field in the Perseus arm is lifted up
  from the plane towards high latitudes.}
 {The magnetic field strength and the electron density we derived for
   the W4 superbubble are important parameters for evolution models of
   superbubbles breaking out of the Galactic plane.}


\maketitle
\section{Introduction}

\begin{table*}[thp]
\caption{Observational parameters}
\label{tab1}
\vspace{-1mm}
\centering
\begin{tabular}{lrrr}
\hline\hline
\multicolumn{1}{c}{Data} &\multicolumn{1}{c}{Urumqi $\lambda$6\ cm}  & \multicolumn{1}{c}{Effelsberg $\lambda$11\ cm} &\multicolumn{1}{c}{EMLS $\lambda$21\ cm}\\
\hline             
Frequency [MHz]                &4800/4963               &2639                         &1400 \\
Bandwidth [MHz]                &600/295                 &80                           &20   \\
HPBW[$\arcmin$]                &9.5                     &4.3                          &9.35 \\
Number of coverages             &6(B)+4(L)               &6(B)+6(L)                    &1(B)+1(L) \\
Integration time [s]           &7.5                     &6                            &2 \\
rms ($I/U,Q$)[mK\ T$_{b}$]      &0.7/0.4                 &2.0/1.5                      &15/8 \\
\vspace{-2mm}\\
\hline
Calibrator                     &3C286                   &3C286                        &3C286 \\
Flux density [Jy]              &7.5                     &11.5                         &14.4 \\
Polarization percentage        &11.3\%                  &9.9\%                        &9.3\% \\
Polarization angle [$\degr$]   &33                      &33                           &32 \\
\hline
\end{tabular}
\vspace{-1mm}
\end{table*}

A large number of superbubbles and supershells have been found and
discussed within our own Galaxy \citep[e.g.][]{Heiles79, Hu81,
  Heiles84, Koo92, Maciejewski96, Uyaniker01, Griffiths00,
  Griffiths06, Pidopryhora07} and in external galaxies
\citep[e.g.][]{Meaburn80, Graham82, Brinks86, Deul90, Kim98}.  They
are created either by energetic stellar winds from OB star
associations, by multiple supernova explosions, or by a
combination. Besides their identification by shell-like \ion{H}{I}
emission, superbubbles and supershells were also traced by soft X-ray
\citep[e.g.][]{Cash80} and H$\alpha$ emission
\citep[e.g.][]{Reynolds79}.  Superbubbles powered by sufficient energy
will break out of the Galactic plane \citep{Tomisaka86, MacLow88,
  MacLow89} and form vertical ``chimneys'' in the interstellar medium
\citep[e.g. the Stockert chimney,][]{Mueller87}.  A general scenario
was proposed by \citet{Norman89} that chimneys play an important role
in the interaction between the Galactic disk and halo by carrying up
mass, energy, momentum, and magnetic flux. Therefore it is of interest
to understand the properties of the superbubbles and their evolution
processes. However, so far only a few superbubbles and chimneys have
been resolved well and the magnetic field strengths measured
\citep[e.g.][]{Vallee93}.

The W4 superbubble in the Perseus arm is one of the few superbubbles
that extend over several degrees and that have been also previously
resolved well in observations.  It was first identified by
\citet{Normandeau96} from \ion{H}{I} observations with the synthesis
telescope of the Dominion Radio Astrophysical Observatory (DRAO)
\citep{Landecker00}. A cone-shaped cavity was revealed to open upwards
from its powering source, the young open star cluster OCl~352, which
includes nine O-type stars. The lower part of the cavity is bounded by
the well-known Galactic \ion{H}{II} region W4. Evidence was found for
an energetic outflow towards high positive latitudes, because
\ion{H}{I} streams were seen to point away from OCl~352.  Because of
the open conical shape viewed in the neutral gas, the entire structure
was first named ``W4 chimney''. Follow-up H$\alpha$ observations
\citep{Dennison97} showed that the ionized gas is detected in the
periphery of the \ion{H}{I} cavity and revealed an enclosed
superbubble rather than a chimney, which is sealed at about $b =
7\degr$. As shown by \citet{West07}, the appearance of this huge
structure seen from the 1.4~GHz radio continuum emission also
resembles a closed bubble. We therefore refer to it as ``W4
superbubble'' in this work.  \citet{Basu99} explained the
morphological difference between a chimney and a superbubble by the
penetration of UV photons to the higher latitude regions. Based on the
study of infrared and radio continuum data, \citet{Terebey03} find a
large leakage of UV photons from OCl~352. This leakage was indirectly
confirmed by the discovery of an even larger high-latitude H$\alpha$
loop on top of the W4 superbubble extending up to $b = 30\degr$
\citep{Reynolds01}.

The distance to the W4 superbubble was taken as 2.2~kpc in the work of
\citet{Normandeau96}, \citet{Dennison97}, \citet{Reynolds01}, and
\citet{Terebey03}, while 2.35~kpc was used by \citet{Basu99} and
\citet{West07}. Through measuring the trigonometric parallax of the
methanol maser, \citet{Xu06} got 1.95~kpc for the distance of the
W3~OH region. Considering a probable association between the
\ion{H}{II} regions W3 and W4 that gives another constraint on the
distance of the W4 superbubble, we adopt the median value of 2.2~kpc
in the following analysis.

The age of the W4 superbubble has not been determined with high
precision. \citet{Dennison97} estimated an age between 6.4 and
9.6~Myr, while \citet{Basu99} derived a lower value of about
2.5~Myr. According to the formation of the large high-latitude
H$\alpha$ loop found on top of the W4 superbubble, \citet{Reynolds01}
argue that the formation of the W4 superbubble is not a single event,
but rather caused by sequential star formation. Age estimates,
however, are inevitably affected by the presence and strength of the
magnetic fields. Numerical simulations have indicated that the
magnetic field has a significant impact on the evolution of
superbubbles \citep[e.g.][]{Tomisaka90, Ferriere91, Tomisaka98}. The
expansion of superbubbles differs along and perpendicular to the field
lines and results in an elongated shape \citep{ Avillez05,
  Stil09}. \citet{Komljenovic99} have suggested that magnetic fields
of a few $\mu$G {\it must} exist in the shells of the W4 superbubble
to maintain its highly collimated shape and prevent its fragmentation
from Rayleigh-Taylor instability.  The strength of the line-of-sight
component of the magnetic fields within the shells of the W4
superbubble was first estimated by \citet{West07} based on the
polarization data collected by the DRAO synthesis telescope. They
determined a field strength between 3~$\mu$G and 5~$\mu$G by analysing
lines through the W4 supperbubble shell separated by 5$\arcmin$. No
short-spacing information was included in their data, which is,
however, required to correctly interpret polarization structures
resulting from Faraday rotation in the interstellar medium
\citep[e.g.][]{Reich06, Sun07, Landecker10}.  We discuss the influence
of polarized large-scale emission on this result in Sect.~4.2.

This work presents a study of the radio spectrum and the magnetic
fields within the shells of the W4 superbubble based on new sensitive
multi-frequency radio continuum and polarization observations
including zero spacings. We describe the data sets we used in Sect.~2
and present all observational results and their analysis in
Sect.~3. In Sect.~4, we use a Faraday screen model to derive the
magnetic field strengths in the W4 superbubble. We discuss and
summarize the results in Sect.~5.

\begin{figure*}[!tbhp]
\centering
\resizebox{0.33\textwidth}{!}{\includegraphics[angle=-90]{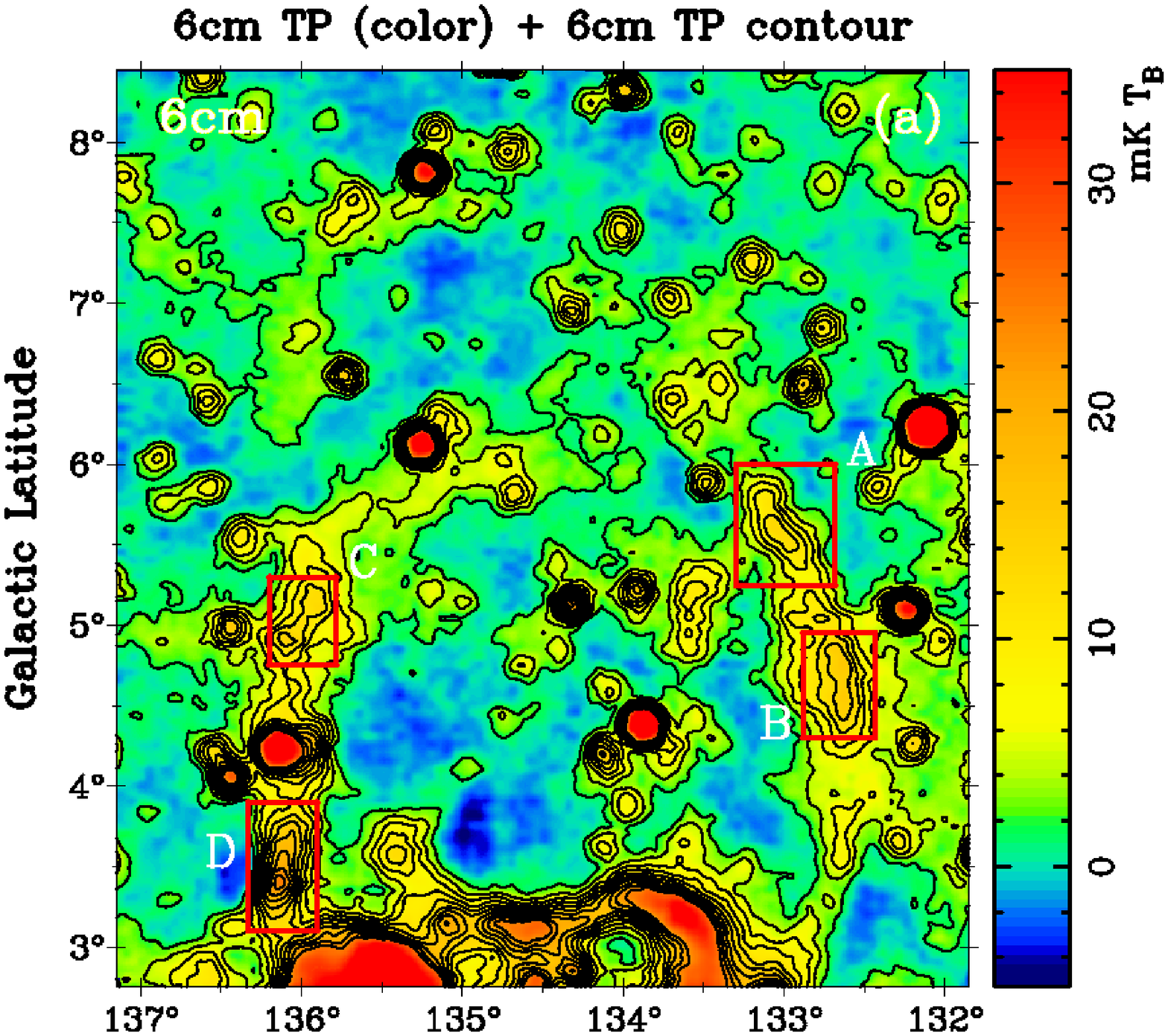}}
\resizebox{0.312\textwidth}{!}{\includegraphics[angle=-90]{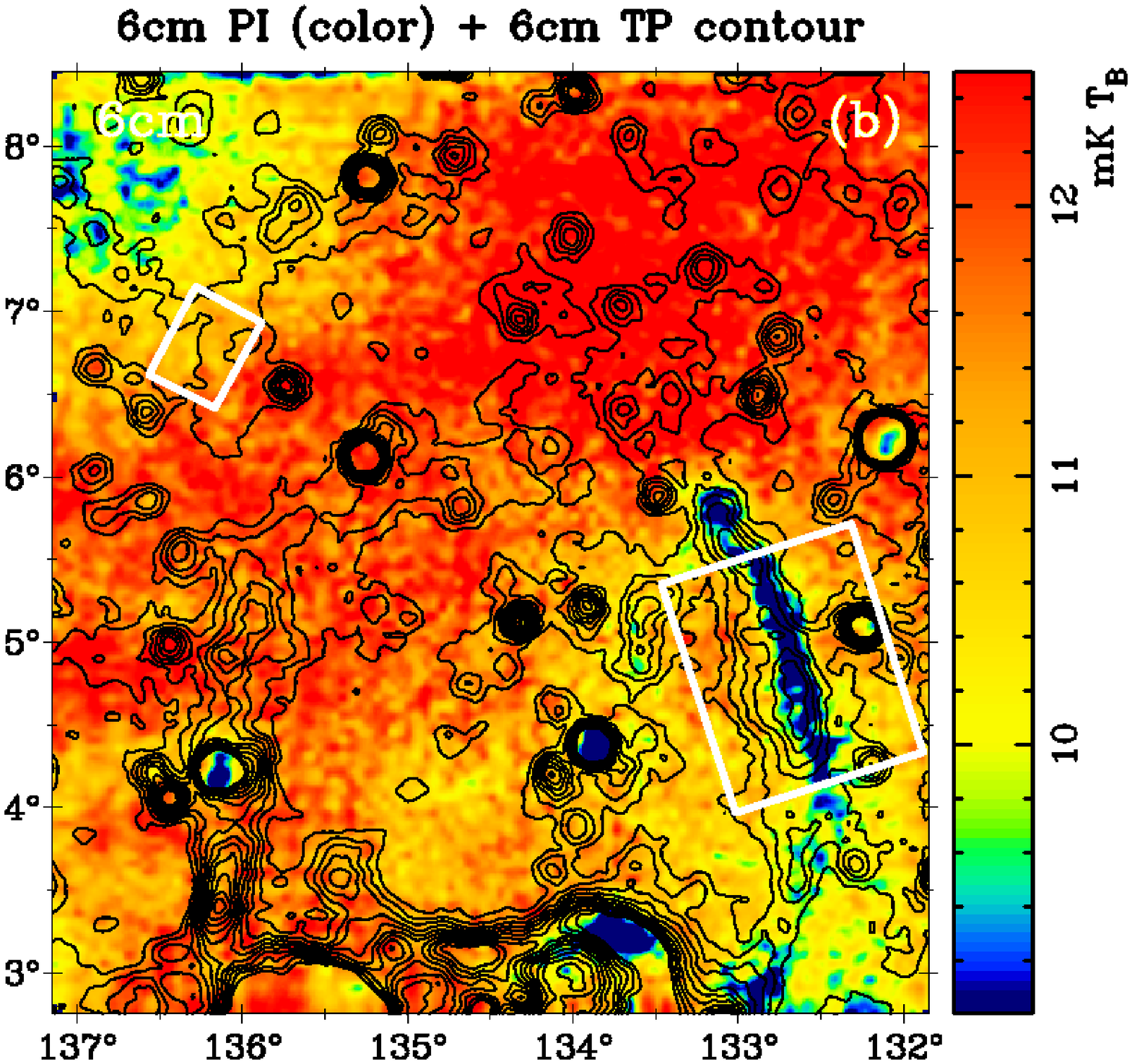}}
\resizebox{0.312\textwidth}{!}{\includegraphics[angle=-90]{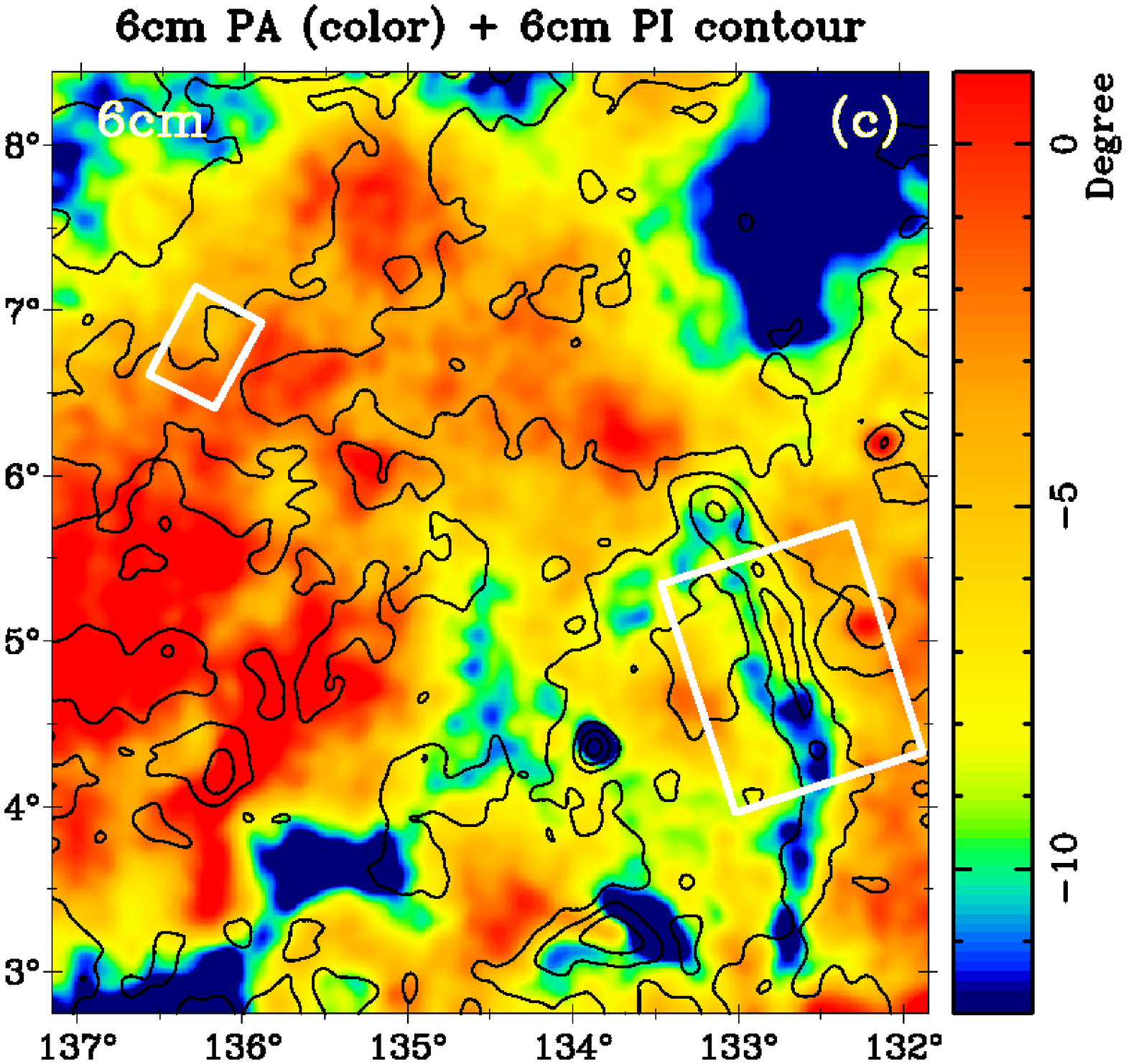}}\\
\vspace{0.2mm}
\resizebox{0.33\textwidth}{!}{\includegraphics[angle=-90]{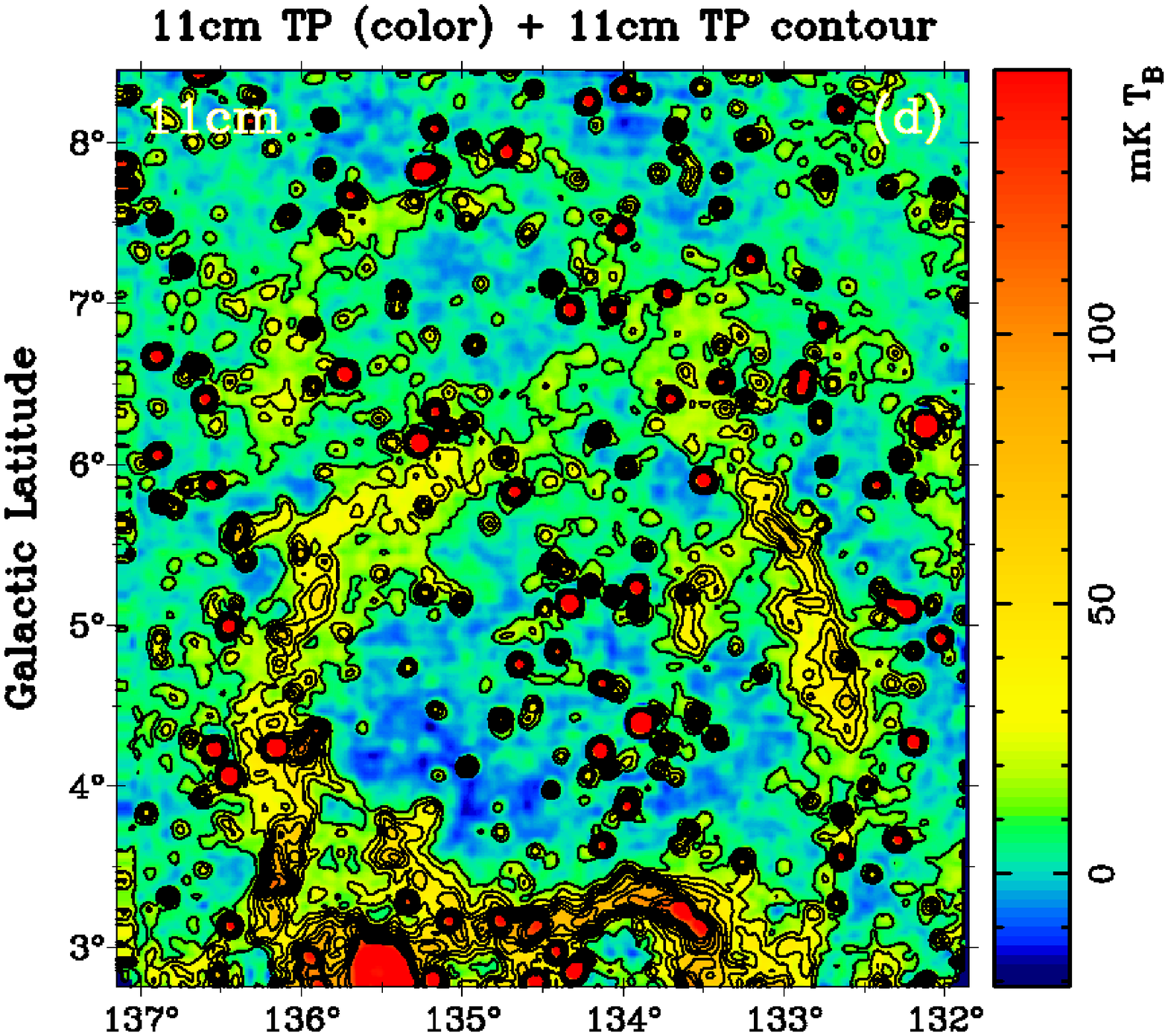}}
\resizebox{0.312\textwidth}{!}{\includegraphics[angle=-90]{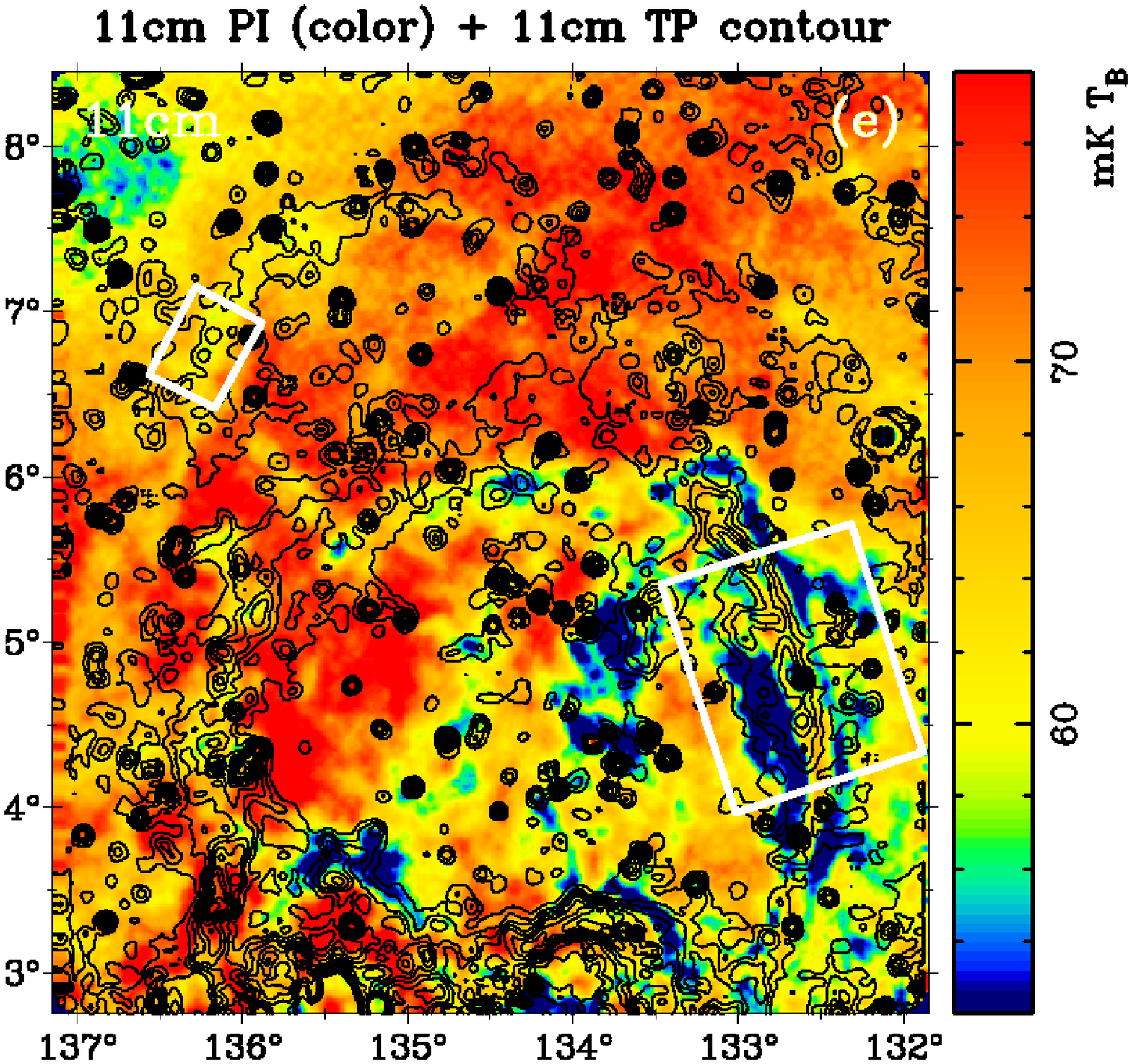}}
\resizebox{0.312\textwidth}{!}{\includegraphics[angle=-90]{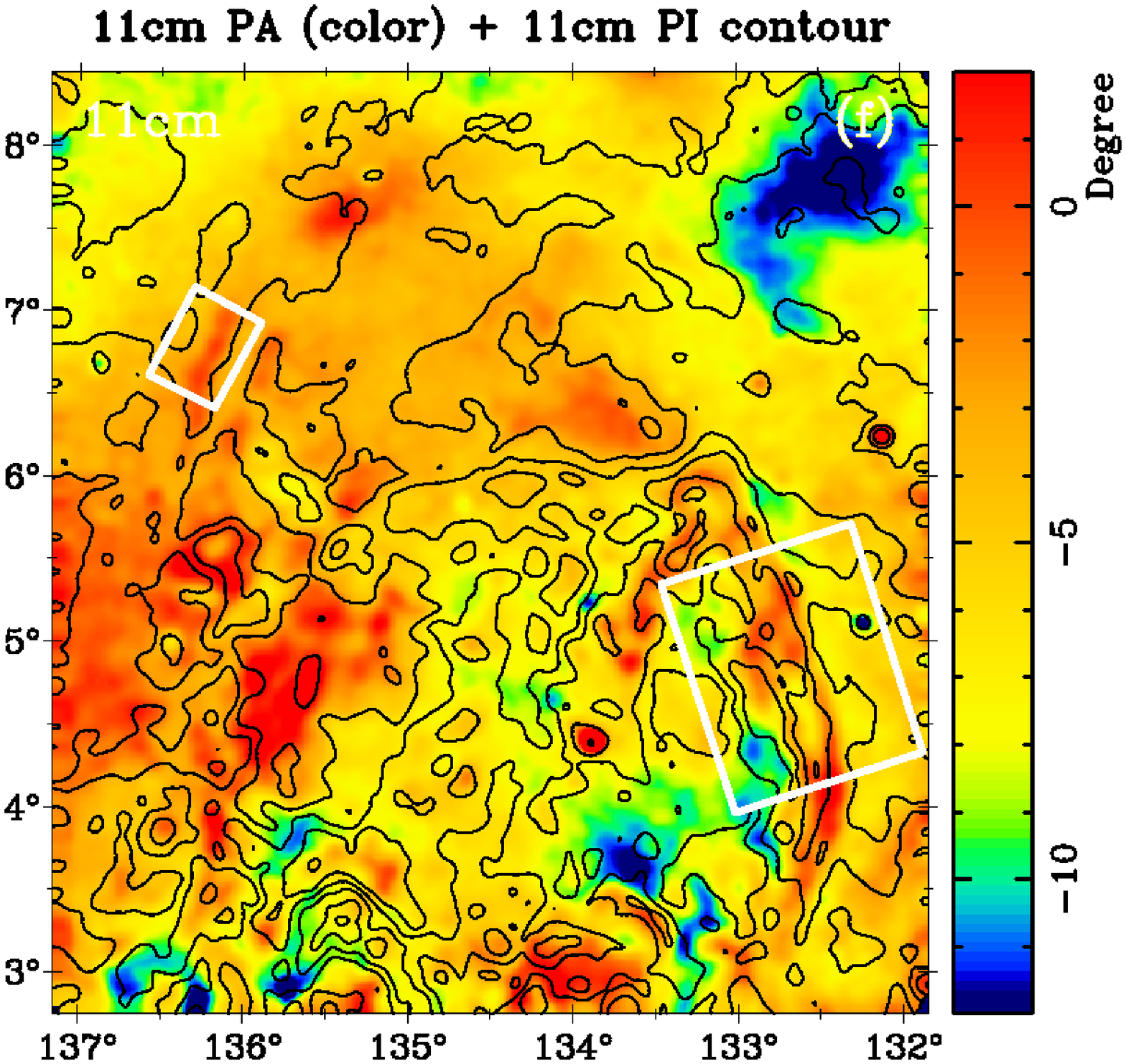}}\\
\vspace{0.2mm}
\resizebox{0.33\textwidth}{!}{\includegraphics[angle=-90]{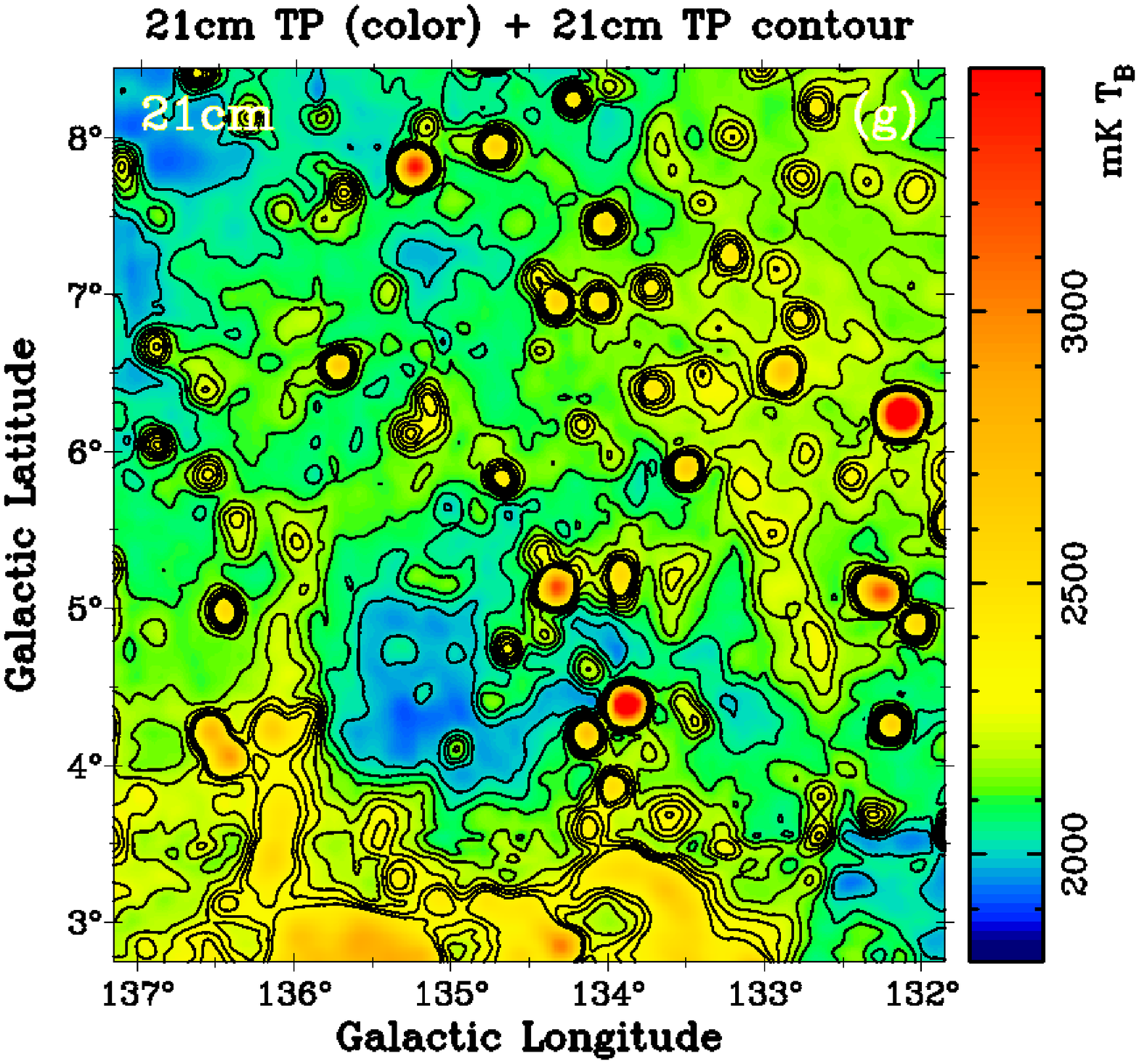}}
\resizebox{0.312\textwidth}{!}{\includegraphics[angle=-90]{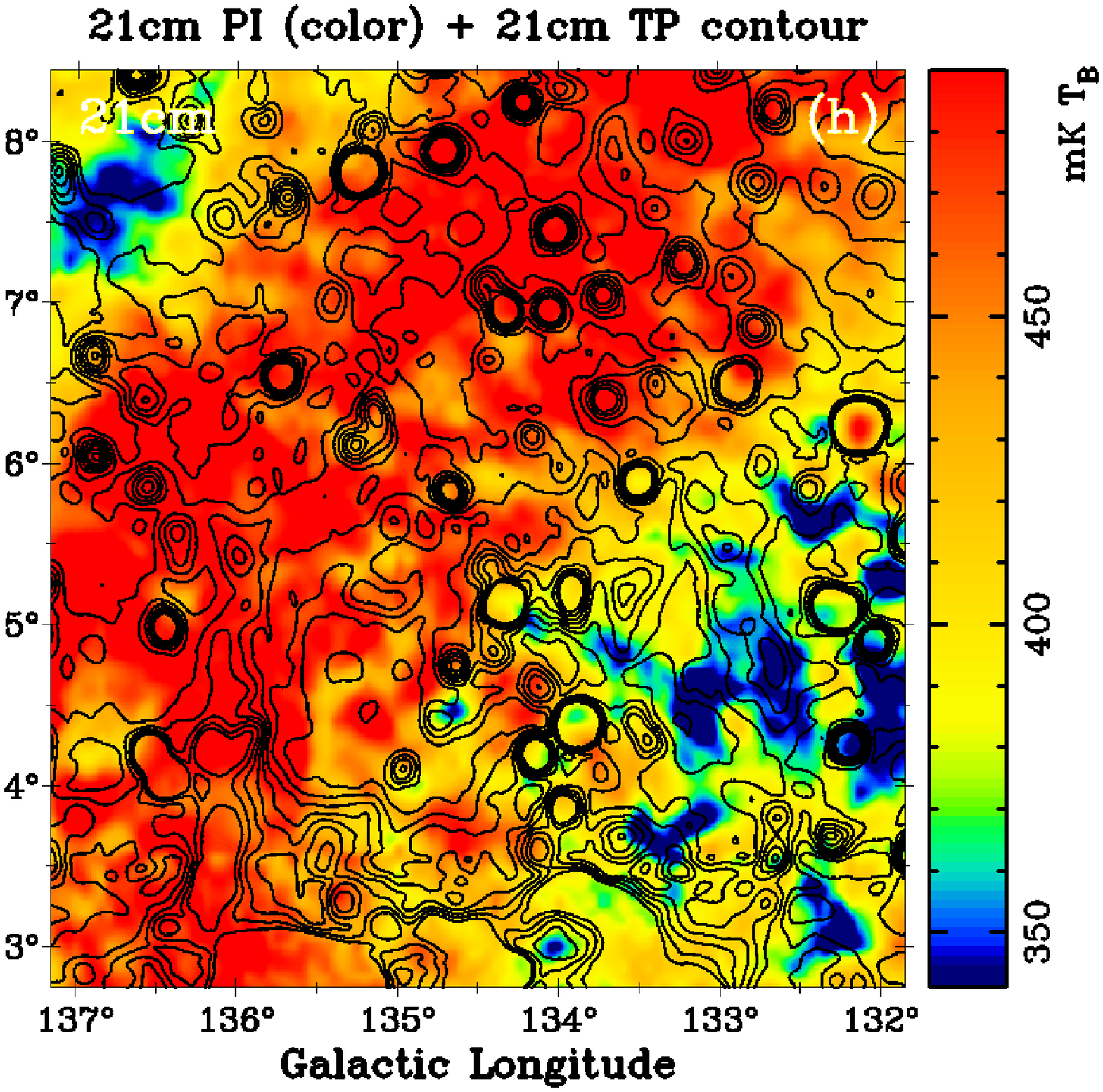}}
\resizebox{0.312\textwidth}{!}{\includegraphics[angle=-90]{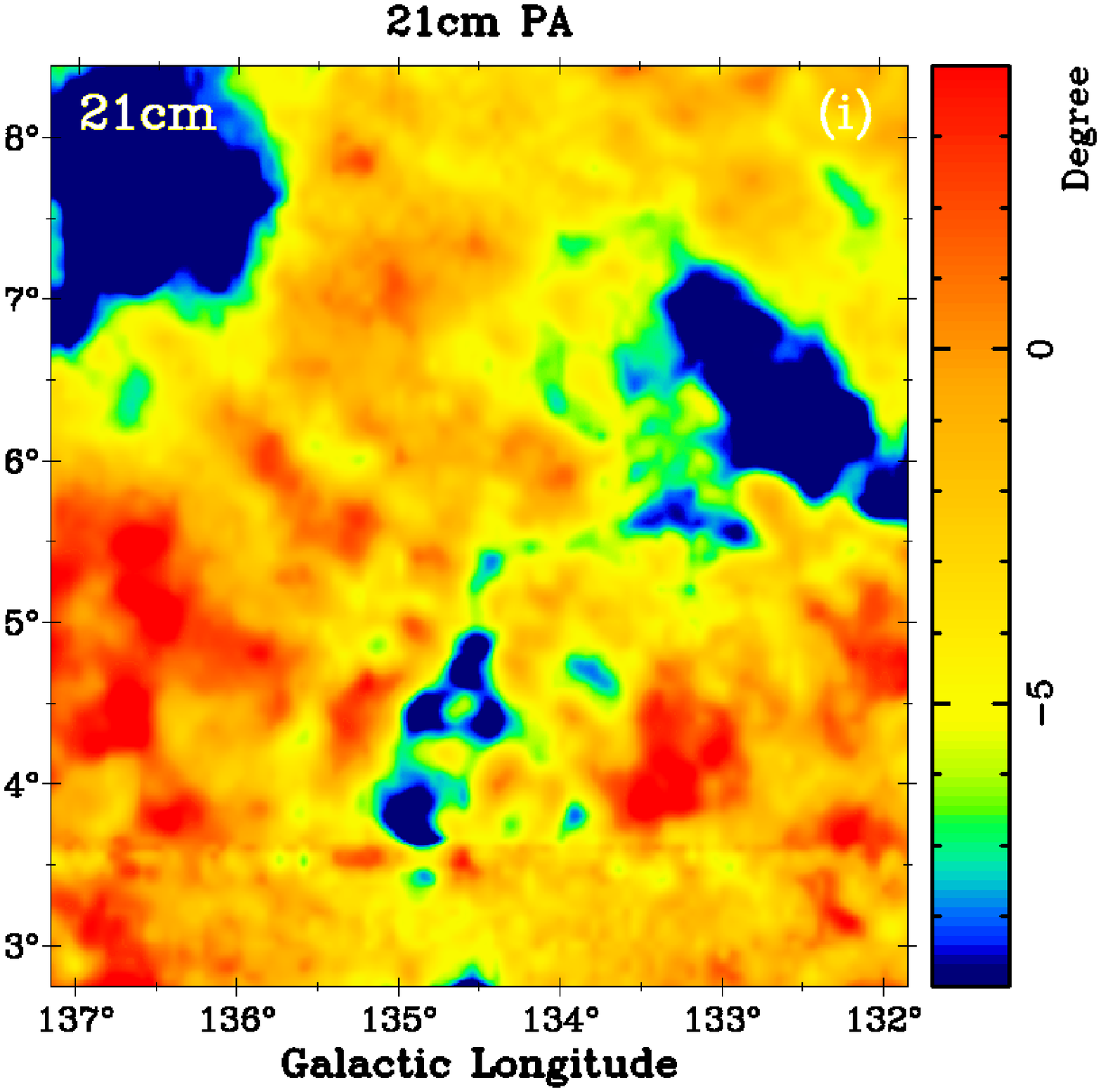}}\\
\caption{Total intensity $I$, polarization intensity $PI$, and
  polarization angle $PA$ images of the W4 superbubble at
  $\lambda$6\ cm ({\it upper panels a, b, c}), $\lambda$11\ cm ({\it
    middle panels d, e, f}), and $\lambda$21\ cm ({\it bottom panels
    g, h, i}). The angular resolutions are 9$\farcm$5 for
  $\lambda$6\ cm images in {\it panels a and b}, 4$\farcm$3 for
  $\lambda$11\ cm images in {\it panels d and e}, and 9$\farcm$35 for
  $\lambda$21\ cm images in {\it panels g, h, and i}. The $PA$ images
  at $\lambda$6\ cm ({\it panel c}) and $\lambda$11\ cm ({\it panel
    f}) have angular resolutions of 12$\arcmin$ and 6$\arcmin$,
  respectively. The total intensity contours for $I$ and $PI$ images
  run from 1.8~mK T$_{b}$ in steps of 2.4~mK T$_{b}$ for
  $\lambda$6\ cm, from 10.0~mK T$_{b}$ in steps of 12.0~mK T$_{b}$ for
  $\lambda$11\ cm (after subtracting strong point-like sources), and
  from 2000~mK T$_{b}$ in steps of 50~mK T$_{b}$ for
  $\lambda$21\ cm. The $PI$ contours on the $PA$ images start from
  5.0~mK T$_{b}$ in steps of 1.0~mK T$_{b}$ for $\lambda$6\ cm, and
  from 50.0~mK T$_{b}$ in steps of 8.0~mK T$_{b}$ for $\lambda$11\ cm.
  $PI$ contours are not overlaid on the $\lambda$21\ cm $PA$ image,
  because no clear correlations can be found. The rectangles in {\it
    panel a} are the regions for TT-plots study, while the rectangles
  in {\it panels b, c, and e, f} indicate the regions used for the
  Faraday screen model fitting.}
\label{ipi}
\end{figure*}

\section{Data}

New sensitive radio continuum and linear polarization observations of
the W4 superbubble have been made at $\lambda$6\ cm with the Urumqi
25-m radio telescope of Xinjiang Astronomical Observatories, Chinese
Academy of Sciences, and at $\lambda$11\ cm with the Effelsberg 100-m
radio telescope of the Max-Planck-Institut f{\"u}r Radioastronomie,
Germany. Total intensity and polarization data at $\lambda$21\ cm were
taken from an unpublished section of the Effelsberg Medium Latitude
Survey (EMLS) \citep{Uyaniker98, Uyaniker99, Reich04}. The Stockert
$\lambda$21\ cm total intensity northern sky survey \citep{Reich82,
  Reich86} and the DRAO 26-m single-dish polarization survey
\citep{Wolleben06} provided the missing large-scale intensities and
the absolute zero levels. We also added high-resolution
$\lambda$21\ cm data observed with the DRAO synthesis telescope by
\citet{West07} and \citet{Landecker10} to the large-scale corrected
EMLS maps.

Technical details of each observation, such as the central frequency,
angular resolution, and sensitivity, are listed in
Table~\ref{tab1}. The bandwidth depolarization
\citep[e.g.][]{Crawford07} for the single-channel observations by the
Urumqi and Effelsberg telescopes was found to be negligible ($<$1\%)
for all the three wavelengths, even for a rotation measure (RM) as
large as 300\ rad~m$^{-2}$.

\subsection{Urumqi $\lambda$6\ cm data}

The $\lambda$6\ cm total intensity and linear polarization data were
obtained between 2007 and 2009 during the conduction of the
Sino-German $\lambda$6\ cm polarization survey of the Galactic plane
\citep{Sun07, Gao10, Sun11a, Xiao11}, using the same observational and
reduction methods. The $\lambda$6\ cm observations had two modes --
the broad-band mode with a central frequency of 4\,800~MHz and a
bandwidth of 600~MHz -- and a narrow-band mode with a central
frequency shifted to 4\,963~MHz and a reduced bandwidth of
295~MHz. Raster scans with a speed of 4$\degr$/min were made along the
Galactic longitude (L) and latitude (B) directions.  A single
observation mostly lasted for about two hours, so that only a portion
of the large target field was covered. Elevations were always high
enough that ground radiation fluctuations can be largely avoided.  Ten
full coverages were processed and then calibrated by 3C286 and 3C295,
the main polarized and un-polarized calibrators. Finally, individual
maps were combined following the procedures described by
\citet{Gao10}. The final image of the W4 superbubble (see
Fig.~\ref{ipi}) is centred at $\ell = 134\fdg5, b=5\fdg6$, covering an
area of $5\fdg3$ $\times$ $5\fdg7$.

\subsection{Effelsberg $\lambda$11\ cm data}

The $\lambda$11\ cm observations were done in the summer of 2008 with
the Effelsberg 100-m telescope. The receiving system was described by
\citet{Uyaniker04}, but was upgraded in 2005 with lower noise
amplifiers and an eight-channel polarimeter.  Here we made use of its
broad-band channel with 80~MHz bandwidth. Data processing and
calibration follow the same procedures as for the $\lambda$6\ cm
observations.

\subsection{Combined $\lambda$21\ cm data}

The $\lambda$21\ cm data were extracted from the EMLS (Reich et al.,
in prep.) with a central frequency of 1\,400~MHz and a bandwidth of
20~MHz. The missing large-scale structures were added by the Stockert
$\lambda$21\ cm survey data \citep{Reich82, Reich86} for total
intensities, and the DRAO 26-m single-dish data \citep{Wolleben06} for
polarization intensities.

We also added the available $\lambda$21\ cm total intensity data from
the Canadian Galactic plane survey \citep{Taylor03}, the polarization
data from \citet{Landecker10}, and the synthesis data from
\citet{West07}, convolved to a 1$\farcm$5 circular beam to the
zero-level restored EMLS maps. A detailed description of the merging
process of polarization data from the DRAO synthesis telescope, the
Effelsberg 100-m and the DRAO 26-m telescopes was given by
\citet{Landecker10}, whose procedure we followed.  The resulting
$\lambda$21\ cm map does not fully cover the area of the
$\lambda$6\ cm, $\lambda$11\ cm, and EMLS $\lambda$21\ cm maps.

\subsection{Absolute zero-level restoration for the $\lambda$6\ cm 
and $\lambda$11\ cm polarization data}

Interferometric data miss short-spacings and single-dish maps are
generally set on arbitrary zero levels. As emphasized by
\citet{Reich06}, missing large-scale structure means that diffuse
polarized emission originating in the magnetized interstellar medium
cannot be properly interpreted.

The observed polarization $U$ and $Q$ maps at $\lambda$6\ cm and
$\lambda$11\ cm were set to zero at their boundaries and thus miss
polarized emission from larger-scale components. The zero-level
problem was solved for the $\lambda$6\ cm polarization data of the
Urumqi survey \citep[e.g.][]{Sun07, Gao10} by adding the missing
large-scale component by extrapolation of the WMAP K-band (22.8~GHz)
polarization data \citep{Page07, Hinshaw09}, which are on absolute
levels.  This is based on the assumption that Faraday rotation of the
Galactic diffuse interstellar medium is negligible for frequencies as
high as 4.8~GHz, which is valid except for the inner Galaxy
\citep{Sun11a}. For the outer region within a few degrees away from
the Galactic plane, such as the W4 superbubble, Faraday rotation
effects can certainly be neglected between $\lambda$6\ cm and the WMAP
K-band.

For the W4 superbubble, we followed the same polarization zero-level
restoration scheme as described by \citet{Gao10}, but using the most
recent WMAP K-band 9-yr data \citep{Bennett13}. First, the Urumqi
$\lambda$6\ cm polarization $U$ and $Q$ maps and the corresponding
WMAP K-band polarization maps were convolved to a common angular
resolution of 2$\degr$. The convolved WMAP $U$ and $Q$ maps needed to
be scaled to $\lambda$6\ cm by a factor of
$(\frac{4800}{22800})^{\beta}$, where ${\beta}$ is the brightness
temperature spectral index (T$_{b} \sim \nu^{\beta}$) for diffuse
polarized emission.  Spectral index maps for polarization intensities
at 75$\arcmin$ resolution were derived from the combined
$\lambda$21\ cm (Effelsberg 100-m and DRAO 26-m) and the WMAP K-band
data, and also for control for the WMAP K- and Ka-band (33~GHz) data
for the entire W4 superbubble area. In both cases, the mean of
${\beta}$ was close to $-$3, i.e. for $\lambda$21\ cm versus K-band,
${\beta}$ = $-$2.97$\pm$0.04, and for K-band versus Ka-band, ${\beta}$
= $-$2.98$\pm$0.31, indicating that depolarization at $\lambda$21\ cm
is not important for large-scale emission. Finally, the
${\beta}$-scaled WMAP $U$ and $Q$ data at 2$\degr$ angular resolution
were compared with the convolved $\lambda$6\ cm $U$ and $Q$ data,
respectively. The differences were taken as the missing large-scale
components and added to the observed $\lambda$6\ cm data at their
original angular resolution.  The $\lambda$6\ cm polarization
intensity ($PI$) was then calculated from the restored $U$ and $Q$ as
$PI = \sqrt{U^2+Q^2}$, and the polarization angle ($PA$) was obtained
by $PA = \frac{1}{2}$atan$({U/Q})$.

The Faraday rotation effect at $\lambda$11\ cm is about 3.4 times that
of $\lambda$6\ cm ($\lambda^{2}$ dependence), so a successful
zero-level recovery using high-frequency WMAP K-band data requires a
careful check on the rotation measure (RM) in the area of
interest. For the W4 superbubble area, we calculated a RM map between
$\lambda$21\ cm (Effelsberg + DRAO) and $\lambda$1.3\ cm (WMAP K-band)
from the polarization angle data: RM = $(PA_{\rm 1.3cm} - PA_{\rm
  21cm}$) / ($0.013^{2} - 0.214^{2})$ at an angular resolution of
75$\arcmin$. The RMs are found to be small in general and the mean is
0.74 $\pm$1.01 rad~m$^{-2}$, which causes only small deviations in
$PA$ of $-$1$\degr$ to 3$\fdg$3 between the Effelsberg $\lambda$11\ cm
and the WMAP $\lambda$1.3\ cm data. Therefore, we restored the missing
large-scale polarized emission for the Effelsberg $\lambda$11\ cm data
by the same procedures as for the Urumqi $\lambda$6\ cm data.

\begin{figure}
\centering
\includegraphics[angle=-90, width=0.45\textwidth]{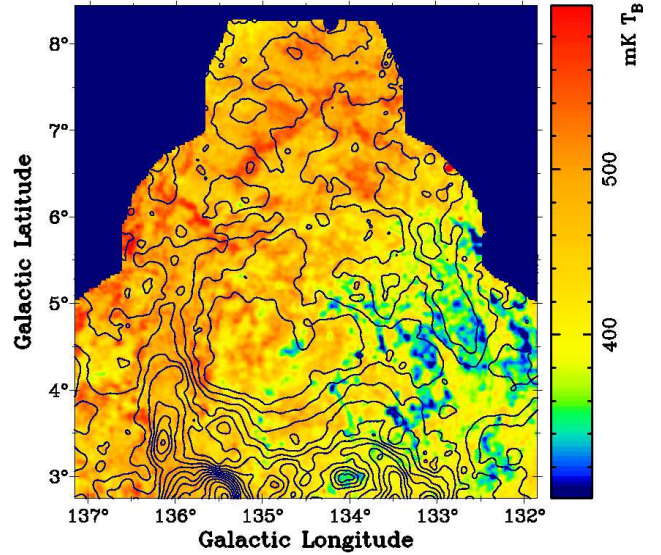}
\caption{Combined $\lambda$21\ cm polarized emission of the W4
  superbubble with data from the DRAO interferometer, the Effelsberg
  100-m, and the DRAO 26-m single-dish telescopes at an angular
  resolution of 3$\arcmin$. Overlaid $\lambda$21\ cm contours are from
  the source subtracted and spatially filtered total intensity map
  (see text). Contour lines run from 2000~mK T$_{b}$ in steps of 50~mK
  T$_{b}$.}
\label{west}
\end{figure}

\begin{figure*}[!tbhp]
\centering
\begin{minipage}[bth]{0.32\textwidth}
\vspace{0.5cm}
\centering
\includegraphics[angle=-90, width=5.1cm]{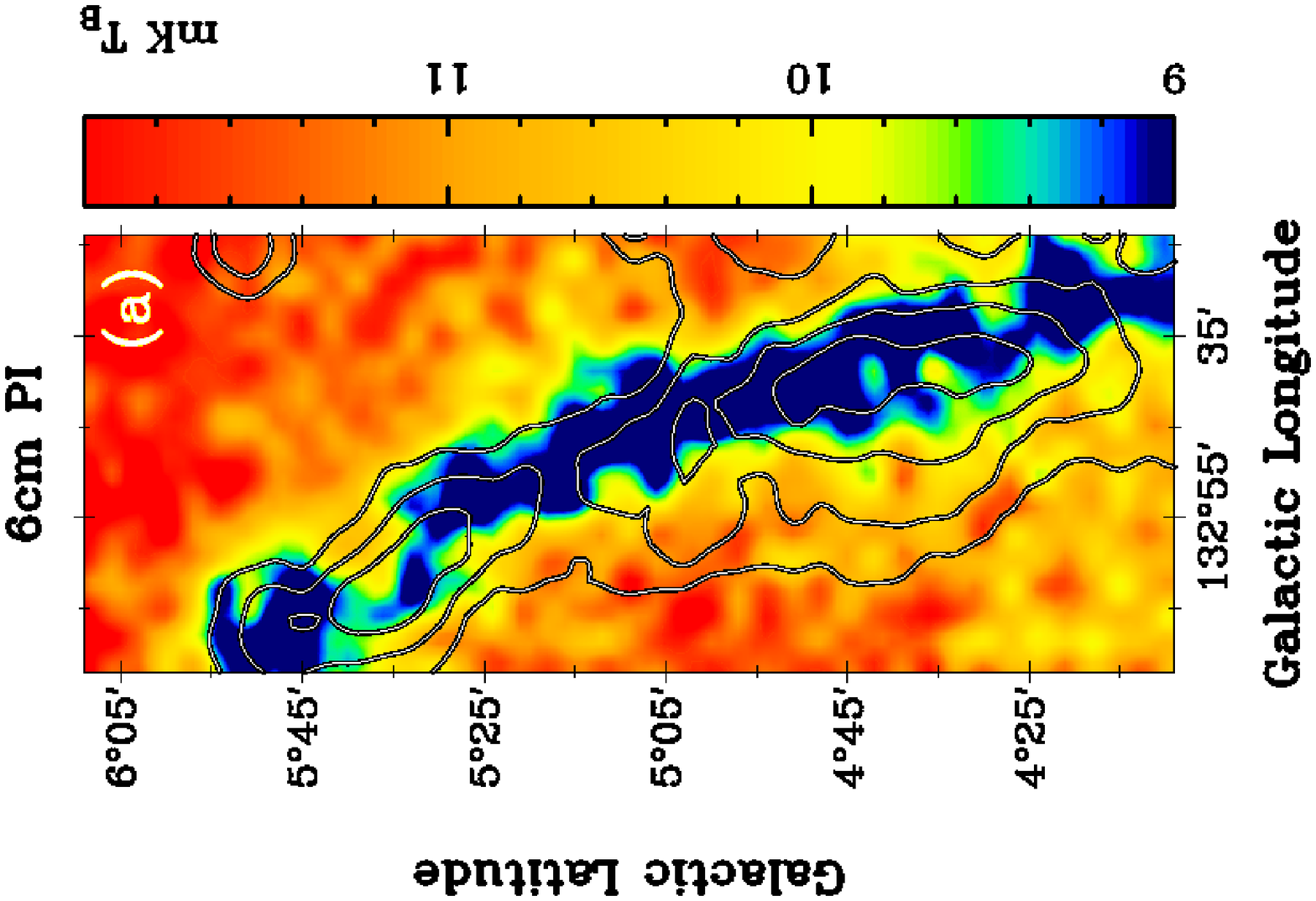}
\end{minipage}
\begin{minipage}[bth]{0.32\textwidth}
\vspace{0.5cm}
\centering
\includegraphics[angle=-90, width=5.1cm]{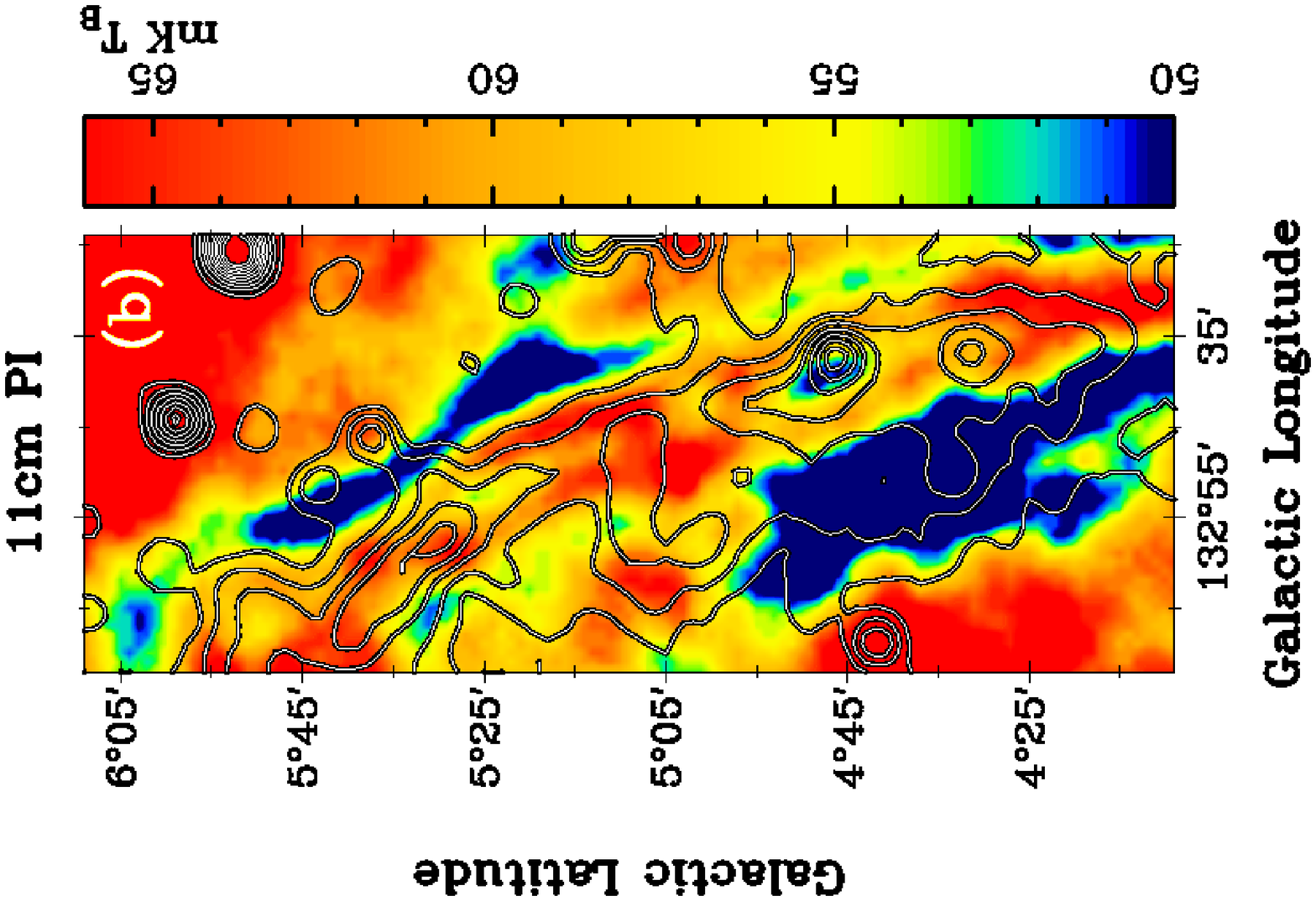}
\end{minipage}
\begin{minipage}[bth]{0.32\textwidth}
\vspace{0.5cm}
\centering
\includegraphics[angle=-90, width=5.1cm]{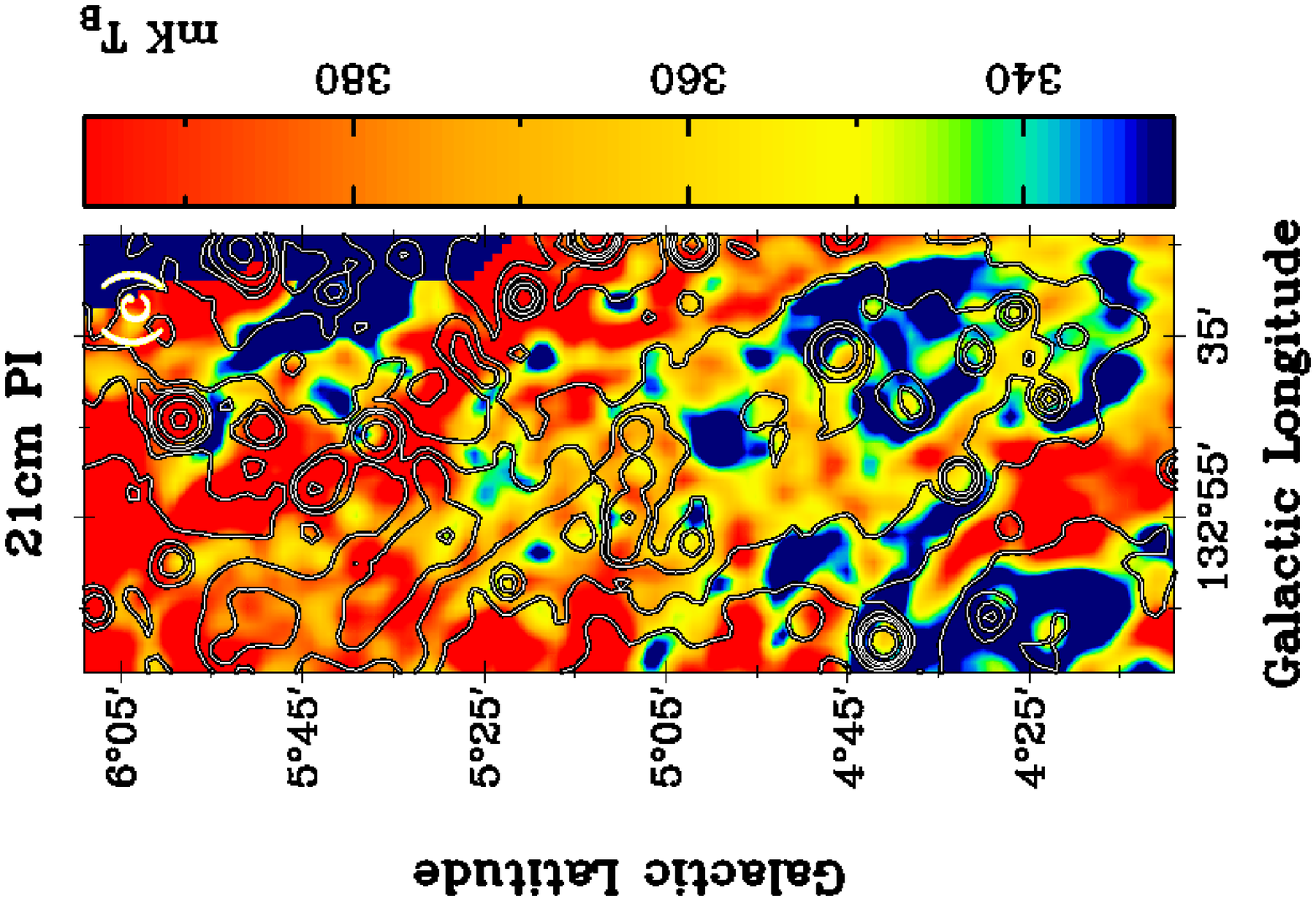}
\end{minipage}
\caption{Depolarization along the western shell of the W4 superbubble
  at $\lambda$6\ cm ({\it panel a}), where total intensity contours
  start at 4.9~mK T$_{b}$ and run in steps of 3.0~mK\ T$_{b}$, and at
  $\lambda$11\ cm ({\it panel b}) where total intensity contours start
  from 9.0~mK\ T$_{b}$ in steps of 13.5~mK\ T$_{b}$, and at
  $\lambda$21\ cm ({\it panel c}) where the total intensity (DRAO
  interferometer + Effelsberg + DRAO 26-m) contours run from 2000 +
  (n-1)~$\times$~50~mK\ T$_{b}$ (n = 1, 2, 3, 4), and 2500 +
  (n-5)~$\times$~800~mK\ T$_{b}$ (n = 5, 6, 7, 8). The angular
  resolutions for the $\lambda$6\ cm, $\lambda$11\ cm, and
  $\lambda$21\ cm images are 9$\farcm$5, 4$\farcm$3, and 3$\arcmin$,
  respectively.}
\label{wshell}
\end{figure*}

\begin{figure*}
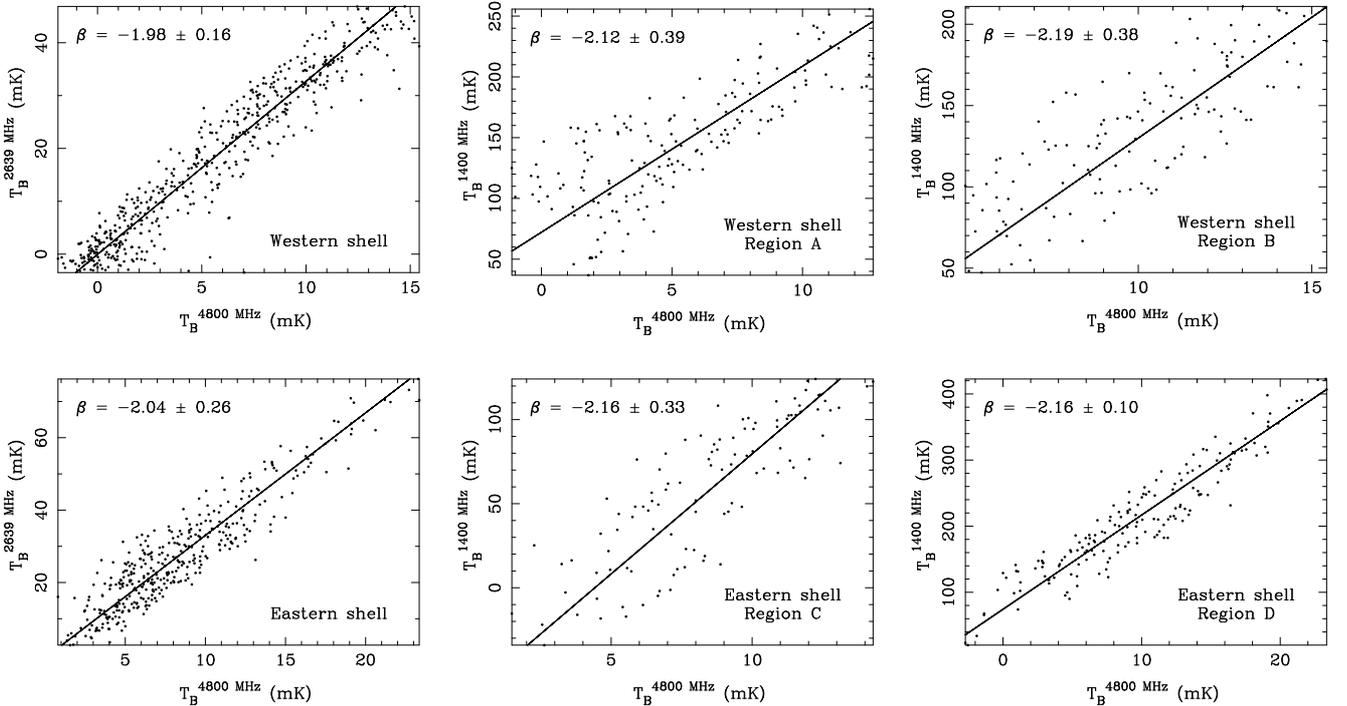
 
\centering
\begin{minipage}[tbh]{0.32\textwidth}
\vspace{0.5cm}
\centering
\includegraphics[angle=-90, width=5.5cm]{24952fg4a.ps}
\end{minipage}
\centering
\begin{minipage}[tbh]{0.32\textwidth}
\vspace{0.5cm}
\centering
\includegraphics[angle=-90, width=5.5cm]{24952fg4b.ps}
\end{minipage}
\centering
\begin{minipage}[tbh]{0.32\textwidth}
\vspace{0.5cm}
\centering
\includegraphics[angle=-90, width=5.5cm]{24952fg4c.ps}
\end{minipage}
\begin{minipage}[tbh]{0.32\textwidth}
\vspace{0.5cm}
\centering
\includegraphics[angle=-90, width=5.5cm]{24952fg4d.ps}
\end{minipage}
\begin{minipage}[tbh]{0.32\textwidth}
\vspace{0.5cm}
\centering
\includegraphics[angle=-90, width=5.5cm]{24952fg4e.ps}
\end{minipage}
\begin{minipage}[tbh]{0.32\textwidth}
\vspace{0.5cm}
\centering
\includegraphics[angle=-90, width=5.5cm]{24952fg4f.ps}
\end{minipage}
\caption{TT-plot results for the western shell ({\it upper panels})
  and the eastern shell ({\it lower panels}) of the W4 superbubble
  between $\lambda$6\ cm and $\lambda$11\ cm ({\it left panels}) and
  for the four regions (A, B, C, and D) in the shells (see
  Fig.~\ref{ipi}) between $\lambda$6\ cm and $\lambda$21\ cm.}
\label{tt}
\end{figure*}

\section{Results}

We present the total intensity $I$, the polarization intensity $PI$ of
the W4 superbubble at $\lambda$6\ cm, $\lambda$11\ cm, and
$\lambda$21\ cm at angular resolutions of 9$\farcm$5, 4$\farcm$3, and
9$\farcm$35 in Fig.~\ref{ipi}, where the $\lambda$21\ cm images are
from the combined EMLS and DRAO 26-m single dish map. The angular
resolutions of the $\lambda$6\ cm and $\lambda$11\ cm $PA$ images are
convolved to 12$\arcmin$ and 6$\arcmin$, respectively, to increase the
signal-to-noise ratio. The $\lambda$21\ cm $PA$ image is shown at its
original resolution of 9$\farcm$35. A higher angular resolution
$\lambda$21\ cm $PI$ map combined from the DRAO synthesis telescope
\citep{West07}, the EMLS, and the DRAO 26-m telescope convolved to a
3$\arcmin$ circular beam is displayed in Fig.~\ref{west}. This map is
overlaid with contours of a source-subtracted and unsharp-masked
$\lambda$21\ cm total intensity map, which was combined from DRAO
synthesis telescope, EMLS, and Stockert survey data.  The singled-out
$PI$ images for the western\footnote{The rotation angle between the
  Galactic plane and the Celestial equator is small ($\sim$ 25$\degr$)
  in the W4 superbubble area, we thus refer to the larger longitudes
  as ``east'', and the smaller longitudes as ``west'' in this work.}
shell of the W4 superbubble at $\lambda$6\ cm, $\lambda$11\ cm, and
$\lambda$21\ cm are shown in Fig.~\ref{wshell}. For all the images,
the $\lambda$6\ cm and $\lambda$11\ cm total intensity images are the
observed ones on relative zero levels, while all the others are
restored to absolute zero levels.

\subsection{Total intensity emission and spectral information}

In all total intensity images, the W4 superbubble shows up by two
bright ridges emerging from the \ion{H}{II} region complex W4 at about
$\ell \sim 133\degr$ and $\ell \sim 136\degr$, extending towards high
latitudes. These two ridges form a loop structure, since they close at
about $b = 6\degr$. From the $\lambda$6\ cm and $\lambda$11\ cm
images, a faint high-latitude extension of the western shell can be
traced to more than $b =7\degr$. Another partial shell-like structure
can be distinguished above the major eastern shell of the W4
superbubble and merges with the western extension at about $b
=8\degr$. According to the apparent size determined by the maximum
lengths on the Galactic longitude and latitude dimensions, $\Delta\ell
\times \Delta b = 3\fdg8 \times 7\fdg1$ ($\Delta b$ is calculated with
respect to the position of OCl~352 at $b = 0\fdg$9), the physical size
of the W4 superbubble is about 150~pc in width and 270~pc in height
for a distance of 2.2~kpc. These results are consistent with previous
estimates \citep[e.g.][]{Dennison97}.

Based on the tight correlation between radio continuum and H$\alpha$
emission and the flat radio spectrum derived between 408~MHz and
1420~MHz, \citet{West07} conclude that the radio continuum emission
from the shells of the W4 superbubble is optically thin thermal
emission. We confirmed this by deriving the spectral index from the
present data using TT-plots \citep{Turtle62} between 1.4~GHz and
4.8~GHz. After subtracting the NVSS sources \citep{Condon98}, the
brightness temperature spectral index between $\lambda$6\ cm and
$\lambda$11\ cm was found to be $\beta = -1.98\pm0.16$ for the western
shell and $\beta = -2.04\pm0.26$ for the eastern shell (Fig.~\ref{tt},
{\it left panels}). Between $\lambda$6\ cm and $\lambda$21\ cm, we
obtained $\beta = -2.12\pm0.39$ for region `A', $\beta = -2.19\pm0.38$
for region `B', and $\beta = -2.16\pm0.33$ and $\beta = -2.16\pm0.10$
for the regions `C' and `D', respectively for the marked regions in
Fig.~\ref{ipi}. These results agree with optical-thin thermal
emission, which we use for the passive Faraday screen model to derive
the magnetic fields within the shells of the W4 superbubble in
Sect.~4.2.

\subsection{Polarization}

Unlike the resemblance in the total intensity images shown for the
three different wavelengths, polarization behaves much differently. At
$\lambda$6\ cm, the most pronounced feature related to the W4
superbubble is the strong depolarization that nicely follows the
western shell and the PA deviation across the shell. No other
prominent and closely related $PI$ and $PA$ features are seen along
any other shell structures.

The $\lambda$11\ cm polarization image discloses more structural
details owing to its higher angular resolution. The polarization
intensities and angles vary strongly in the western shell area of the
W4 superbubble (see Fig.~\ref{wshell}). Other than at $\lambda$6\ cm,
depolarization is also seen for some sections of the eastern shell and
the high-latitude filament centred approximately at $\ell = 136\degr,
b = 7\degr$. This, in conjunction with the $\lambda$6\ cm $PI$ image,
implies that RM varies in values and regularity over the W4
superbubble and is smaller in the eastern shell than in the western
shell. Polarization structures are also visible in the bubble's
interior.  A notable feature is the arc-like structure running from
longitude $133\degr$ to $134\degr$ and latitude $6\fdg0$ to $4\fdg5$,
which is also visible but fainter at $\lambda$6\ cm.

At $\lambda$21\ cm, no strong structural coincidence can be identified
between total intensity and polarization at first
glance. \citet{West07} saw the western and eastern shells of the W4
superbubble in depolarization in their $\lambda$21\ cm map using DRAO
synthesis data alone.  When averaging their data for a 10$\arcmin$
wide latitude stripe centred at $b = 4\fdg5$, the polarization
intensity for the eastern shell drops from a surrounding level of
about 65~mK T$_{b}$ to 40~mK T$_{b}$, and for the western shell it
decreases from 50~mK T$_{b}$ to about 40~mK T$_{b}$.  When combining
the DRAO synthesis telescope data with the Effelsberg and DRAO
single-dish data (Fig.~\ref{west}), we found that this depolarization
is much less pronounced than the pure DRAO synthesis telescope
data. The reason is the much higher absolute polarization level of
about 500~mK T$_{b}$ for the eastern and 350~mK T$_{b}$ for the
western shell. Fluctuations of the Galactic polarized emission seem to
mask the depolarization along the W4 superbubble shells.  For a zoomed
view of the western shell at 3$\arcmin$ resolution
(Fig.~\ref{wshell}), there is, however, a morphological coincidence
between total and polarization intensity, indicating that related
depolarization exists.

On larger scales, an inclined broad stripe of strong polarized
emission runs across the image from the lower left (south-east) to the
upper right (north-west) corner, while depolarization patches are seen
in the orthogonal areas of the image.  Both the Effelsberg and the
DRAO single-dish data independently show this feature at
$\lambda$21\ cm. This general distribution is also seen in the
$\lambda$6\ cm and $\lambda$11\ cm polarization data that are
corrected by large-scale emission based on the WMAP K-band data (see
Fig.~\ref{ipi}).

\section{The magnetic field strength in the W4 superbubble shell}

Regular changes in sign or values of RMs of extragalactic sources
across a finite area often indicate the existence of a Galactic
Faraday screen. Sufficient RM sources in a sky area can be used to
determine the line-of-sight component of the magnetic fields within
such foreground screens \citep[e.g.][]{Harvey-Smith11}. There are 40
sources with determined RMs within the W4 superbubble area in the
catalogue of
\citet{Xu14}\footnote{http://zmtt.bao.ac.cn/RM/index.html}
supplemented by the NVSS RM catalogue \citep{Taylor09}.  Most of these
RMs show negative signs, resulting from the orientation of the
Galactic magnetic field running clockwise (viewed from the north
Galactic pole) in this part of the Perseus arm \citep{Han06}. Five RMs
are positive, which might indicate small-scale variations or
source-intrinsic properties.  However, the present RM number density
is too sparse to estimate the magnetic field for shell regions of the
W4 superbubble.

\subsection{The Faraday screen model}

\begin{figure} 
\centering
\includegraphics[angle=0, width=0.5\textwidth]{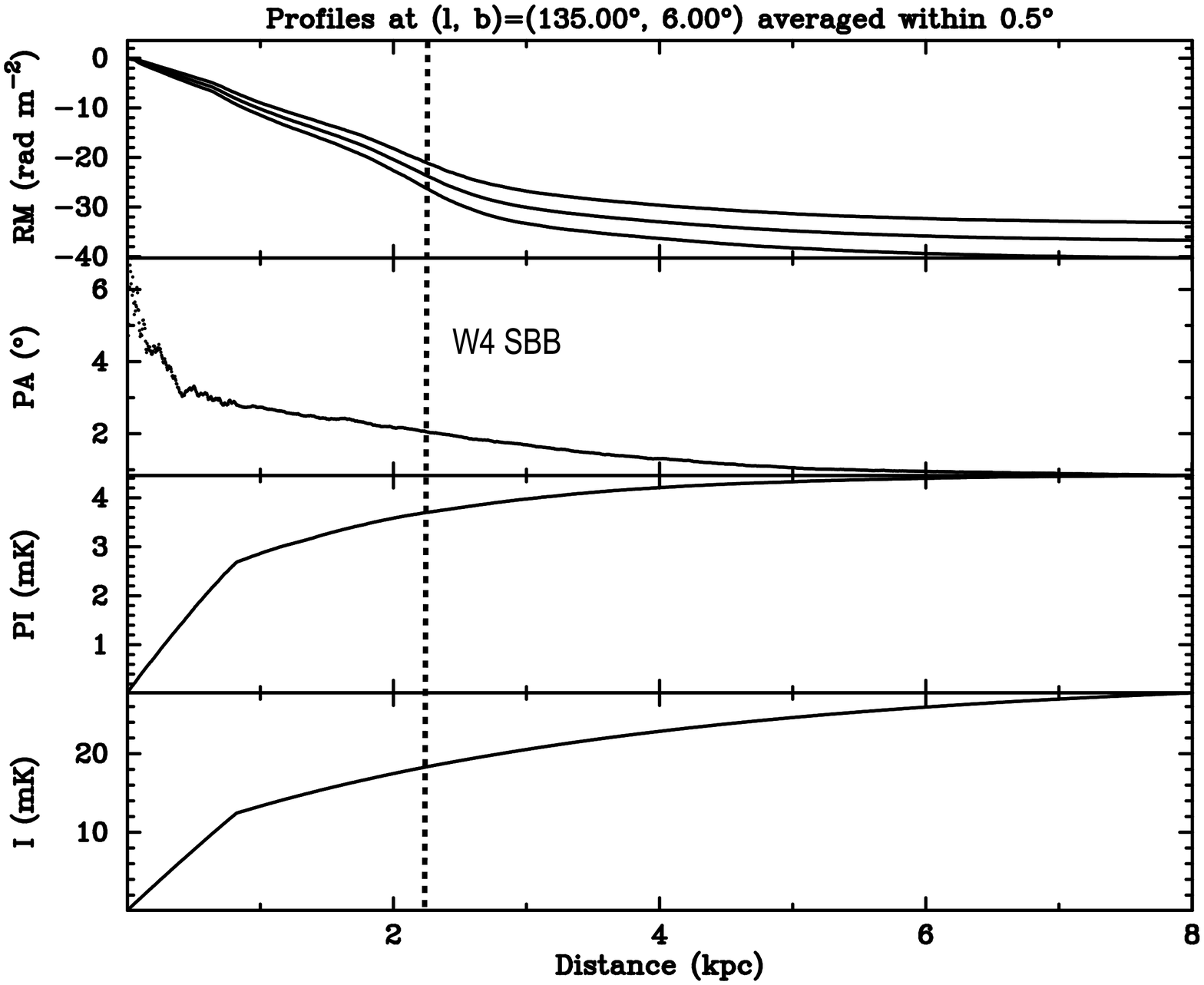}
\caption{Simulations in the direction $\ell = 135\degr, b = 6\degr$
  showing accumulated and spatial averaged RM (including one $\sigma$
  errors), $PA$, $PI$, and $I$ values at $\lambda$6\ cm as a function
  of distance from the Sun.}
\label{sc}
\end{figure}

A passive Faraday screen is a magneto-ionized bubble, filament, or
sheet that does not emit polarized emission, but modulates background
polarized emission passing through it via the Faraday effect. A
successful model for deriving the physical parameters of these Faraday
screens was developed by \citet{Sun07} for screens detected in the
Urumqi $\lambda$6\ cm survey. It was also applied in several follow-up
studies \citep[e.g.][]{Gao10, Xiao11}. In its simplest form, the model
takes the observational fact that the foreground and background
Galactic magnetic field of the screens have the same orientation and
$PA \sim 0\degr$, i.e. parallel to the Galactic plane at high enough
frequencies. We have verified in Sect.~3.1 that the W4 superbubble
exclusively emits optically thin thermal emission. We now check
whether foreground and background $PA$s around the W4 superbubble area
meet the condition of $PA$ $\sim$ 0$\degr$. Starlight polarization
observations of the W4 superbubble area \citep{Heiles00} support the
idea that the Galactic magnetic fields are aligned to the Galactic
plane ($PA = -1\degr\pm7\degr$) for a distance range from 30~pc to
2.85~kpc. This is further supported by the low-frequency polarization
measurements for the large excessive ``Fan'' region
\citep{Spoelstra84}, covering the W4 superbubble area, and also by the
WMAP K- and Ka-band polarization data, where Galactic RMs in this part
of the Galaxy are small, and the large-scale Galactic magnetic fields
\citep{Han06} will only marginally change $PA$s from their intrinsic
values. In addition, we also carried out $\lambda$6\ cm simulations in
the direction of the W4 superbubble using the \citet{Sun08} and
\citet{Sun10} Galactic 3D-emission model, which is based on all-sky
surveys, RMs, and thermal emission data and also includes realistic
magnetic field turbulence \citep{Sun09}. The simulation result is
shown in Fig.~\ref{sc}. Foreground $PA$ of the W4 superbubble at
2.2~kpc is close to zero and differs by just about 1$\degr$ from the
$PA$ of the background emission at distances of 8~kpc or more. We
repeated the simulation for various directions within the W4
superbubble area at $\lambda$6\ cm, and obtained very similar
results. Simulations at $\lambda$11\ cm and $\lambda$21\ cm show very
small $PA$ differences as well. We concluded that the Galactic
3D-emission simulations also support that the assumptions required for
the passive Faraday screen model are satisfied.

The passive Faraday screen model uses three parameters: a
depolarization factor $f$, which describes the reduction of the
background polarization by the Faraday screen; the Faraday screen
foreground polarized emission fraction $c$ with respect to the
observed polarized emission at an off-position away from the screen;
and the polarization rotation angle within the screen $\psi_{s}$,
which is the parameter for calculating RM (RM =
$\psi_{s}/\lambda^{2}$), and the line-of-sight component of the
magnetic field (see Eq.~6). These three parameters fit two
observables: (1) the ratio $PI_{\rm on}/PI_{\rm off}$ and (2) the
angle difference $PA_{\rm on} - PA_{\rm off}$, where ``on'' and
``off'' denote that the line of sight passes through (on) or not (off)
the screen. The Faraday screen model equation is given in
Eq.~\ref{eq1} \citep[see][for a detailed derivation]{Sun07}:

\begin{equation}
\centering
\displaystyle{
\left\{
\begin{array}{cc}
\displaystyle
\frac{PI_{\mathrm{on}}}{PI_{\mathrm{off}}}=\sqrt{\mathit{f}^2(1-c)^2+c^2+2\mathit{f}c(1-c)\cos2\psi_s}\ , \\ \displaystyle
PA_{\mathrm{on}} - PA_{\mathrm{off}}=\frac{1}{2}\arctan\left(\frac{\mathit{f}(1-c)\sin2\psi_s}{c+\mathit{f}(1-c)\cos2\psi_s}\right).
 &
\end{array}
\right.
}
\label{eq1}
\end{equation}

\noindent In the following we allow $f$ to be in the range [0.00,
  1.00] and c between 0.6 to 1.0, when taking the simulation result
(Fig.~\ref{sc}) into account. These parameters vary in steps of 0.01.
$\psi_{s}$ is searched for in the range [$-$90$\degr$, 90$\degr$] in
steps of 1$\degr$.

\begin{figure}
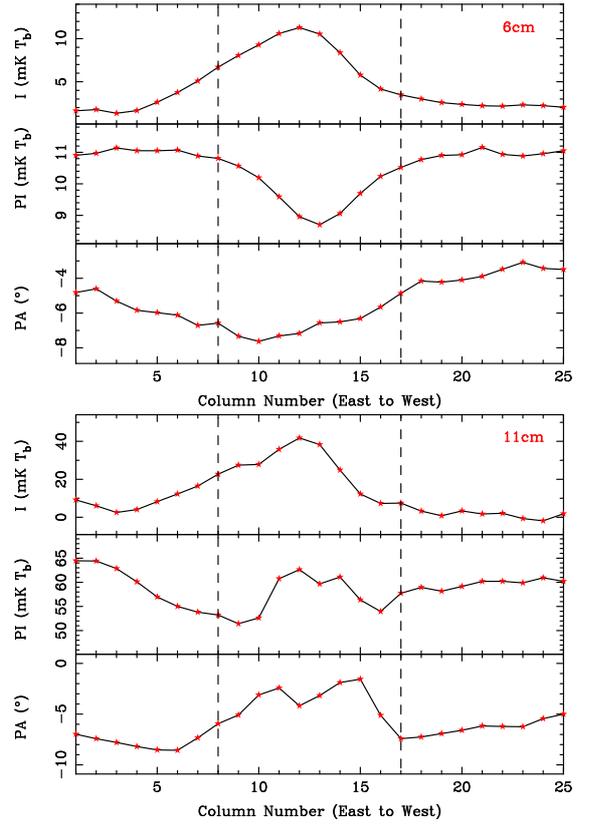
 
\centering
\includegraphics[angle=-90, width=0.4\textwidth]{24952fg6a.ps}
\includegraphics[angle=-90, width=0.4\textwidth]{24952fg6b.ps}
\caption{{\it Upper panel}: profiles of the $\lambda$6\ cm $I$
  (relative zero level), $PI$ (absolute zero level), and $PA$
  (absolute zero level) averaged perpendicular to the western shell of
  the W4 superbubble within the area marked by the large white
  rectangular region in Fig.~\ref{ipi}. The pixel size is
  3$\arcmin$. The vertical dashed lines indicate the ``on - off''
  boundaries of the Faraday screen. {\it Lower panel}: the same as the
  {\it upper panel}, but for the Effelsberg $\lambda$11\ cm data.}
\label{profile}
\end{figure}

\subsection{Analysis of the western shell}

The clearest Faraday effect for the W4 superbubble is the depolarized
western shell seen at $\lambda$6\ cm and $\lambda$11\ cm. With the
measurements at these two wavelengths, we estimated the line-of-sight
component of the magnetic field $B_{//}$ based on the Faraday screen
model described above for a section as indicated by the large white
rectangle in Fig.~\ref{ipi}. Averaged $\lambda$6\ cm and
$\lambda$11\ cm $I$, $PI$, and $PA$ profiles perpendicular to the
western shell were calculated and shown in Fig.~\ref{profile}. The
shell direction is inclined, and the averaged stripes or columns run
parallel to it.  We labelled the result by the column number from east
to west across the western shell (see Fig.~\ref{profile}). The typical
character of a passive Faraday screen is seen at $\lambda$6\ cm; i.e.,
an apparent $PI$ depression is identified with its minimum close to
the peak position of the total intensity $I$.  The very small $PA$
difference between ``on'' and ``off'' positions implies a large
foreground fraction $c$ of the polarized emission.

To calculate the two observables, $PI_{\rm on}/PI_{\rm off}$ and
$PA_{\rm on} - PA_{\rm off}$, the ``on'' and ``off'' positions of the
Faraday screen must be determined first. Combining the profiles of the
$\lambda$6\ cm and $\lambda$11\ cm data, we set the boundary of ``on''
and ``off'' positions of the Faraday screen as marked in
Fig.~\ref{profile}. The column size of 3$\arcmin$ corresponds to
1.9~pc for a distance of 2.2~kpc of the W4 superbubble.  The
$\lambda$6\ cm $PI_{\rm off}$ and $PA_{\rm off}$ were calculated as
the average values of $PI$s and $PA$s on the ``off'' positions (Column
1 - 8 and Column 17 - 25, see Fig.~\ref{profile} {\it upper
  panel}). The central eight (9th to 16th, Fig.~\ref{profile} and
Table 2) columns represent the interior of the Faraday screen and were
used for model fitting.

\begin{sidewaystable*}[p]
\vspace{150mm}
\caption{Faraday screen parameters for the western shell of the W4
  superbubble. Column numbers refer to Fig.~\ref{profile}. The best
  fit ($f$, $c$, $\psi_{s}$) solution has the least deviations from
  the $\lambda$6\ cm and $\lambda$11\ cm observations. We tabulate
  n$_{e}$ (1$\sigma$ error = 5\%) and the magnetic field component
  B$_{//}$ for filling factor $f_{n_e}$ = 0.4 and 1.0. We also list
  n$_{e}$ and B$_{//}$ in case the 12th column represents the longest
  path length (see text).}
\label{t2}
\centering
\renewcommand{\footnoterule}{}
\begin{tabular}{crrrrrrrr}
\hline\hline
\multicolumn{1}{c}{Column No.} &\multicolumn{1}{c}{9} & \multicolumn{1}{c}{10} &\multicolumn{1}{c}{11} &\multicolumn{1}{c}{12} &\multicolumn{1}{c}{13} &\multicolumn{1}{c}{14} &\multicolumn{1}{c}{15} &\multicolumn{1}{c}{16}\\
\hline
($l$, $b$)   &132$\fdg$86, 4$\fdg$76  &132$\fdg$81, 4$\fdg$78  &132$\fdg$76, 4$\fdg$79  &132$\fdg$71, 4$\fdg$81  &132$\fdg$67, 4$\fdg$82   &132$\fdg$62, 4$\fdg$84   &132$\fdg$57, 4$\fdg$86   &132$\fdg$52, 4$\fdg$87        \\      
best fit (f,c,$\psi_s$)  & (0.95, 0.85, $-$19$\degr$)  & (0.78, 0.85, $-$25$\degr$)  & (0.95, 0.91, $-$59$\degr$)  & (0.68, 0.87, $-$64$\degr$)  & (0.51, 0.85, $-$69$\degr$) 
                       & (0.61, 0.88, $-$68$\degr$)  & (0.39, 0.86, $-$28$\degr$)  & (0.59, 0.85, $-$13$\degr$) \\
f                 &0.77$\pm$0.15  &0.77$\pm$0.14 &0.70$\pm$0.19  &0.80$\pm$0.13  &0.54$\pm$0.28  &0.49$\pm$0.28  &0.26$\pm$0.19  &0.82$\pm$0.12  \\

c                 &0.86$\pm$0.02  &0.87$\pm$0.02 &0.90$\pm$0.02  &0.88$\pm$0.02  &0.87$\pm$0.02  &0.88$\pm$0.02  &0.88$\pm$0.02  &0.86$\pm$0.01  \\

$\psi_{s}$ (deg)   &$-$18$\pm$3    &$-$20$\pm$3   &$-$64$\pm$4    &$-$64$\pm$3   &$-$66$\pm$6     &$-$65$\pm$8    &$-$19$\pm$10   &$-$16$\pm$2  \\

RM (rad m$^{-2}$)  &$-$80$\pm$13   &$-$89$\pm$13   &$-$286$\pm$18   &$-$286$\pm$13   &$-$295$\pm$27   &$-$290$\pm$36   &$-$85$\pm$45   &$-$71$\pm$9 \\
\vspace{-2mm}\\
\hline
6cm {\it I$_{obs}$} (mK T$_{b}$)  &6.2           &7.4     &8.7      &9.4     &8.6     &6.5      &3.9      &2.3 \\
{\it EM} (pc cm$^{-6}$)          &47            &56      &66       &71      &65       &49       &30       &17 \\
path length (pc)                &25            &27      &31       &37      &61       &52       &40       &23 \\

\vspace{1mm}\\
$f_{n_e}$ = 0.4                 & & & & & & & & \\
{\it n$_{e}$} (cm$^{-3}$)        &2.2            &2.3            &2.3               &2.2                &1.6     &1.5     &1.4     &1.4 \\
{\it B$_{//}$} ($\mu$G)         &$-$4.6$\pm$0.8  &$-$4.5$\pm$0.7  &$-$12.3$\pm$1.0   &$-$10.9$\pm$0.7    &$-$9.1$\pm$1.0     &$-$11.2$\pm$1.5    &$-$4.8$\pm$2.5    &$-$7.0$\pm$1.0 \\
\vspace{0.5mm}\\
$f_{n_e}$ = 1.0                 & & & & & & & & \\
{\it n$_{e}$} (cm$^{-3}$)        &1.4            &1.4            &1.5               &1.4                &1.0     &1.0     &0.9     &0.9 \\
{\it B$_{//}$} ($\mu$G)         &$-$2.9$\pm$0.5  &$-$2.8$\pm$0.4  &$-$7.8$\pm$0.6   &$-$6.9$\pm$0.5    &$-$5.8$\pm$0.6     &$-$7.1$\pm$1.0    &$-$3.0$\pm$1.6    &$-$4.4$\pm$0.6 \\

\vspace{1mm}\\

\hline
if the 12th column is taken as  & & & & & & & & \\
the longest path length        & & & & & & & & \\
\vspace{0.3mm}\\
$f_{n_e}$ = 0.4                 & & & & & & & &  \\
{\it n$_{e}$} (cm$^{-3}$)    &1.9       &1.9      &1.9      &1.6      &1.6      &1.5      &1.4      &1.4 \\
{\it B$_{//}$} ($\mu$G)     &$-$3.9$\pm$0.7  &$-$3.8$\pm$0.6   &$-$10.4$\pm$0.9   &$-$8.0$\pm$0.5   &$-$9.1$\pm$1.0   &$-$11.2$\pm$1.5   &$-$4.8$\pm$2.5  &$-$7.0$\pm$1.0\\

\vspace{0.5mm}\\
$f_{n_e}$ = 1.0                 & & & & & & & &  \\
{\it n$_{e}$} (cm$^{-3}$)    &1.2       &1.2      &1.2      &1.0      &1.0      &1.0      &0.9      &0.9 \\
{\it B$_{//}$} ($\mu$G)     &$-$2.5$\pm$0.4  &$-$2.4$\pm$0.4   &$-$6.6$\pm$0.5   &$-$5.1$\pm$0.3   &$-$5.8$\pm$0.6   &$-$7.1$\pm$1.0   &$-$3.0$\pm$1.6  &$-$4.4$\pm$0.6\\

\hline
\vspace{1mm}
\end{tabular}\\
{\vspace{-1mm}}
\end{sidewaystable*}

We assumed 1$\sigma$ observational uncertainty for the $\lambda$6\ cm
observations and used Eq.~1 for the Faraday screen model fitting. A
series of ($f$, $c$, $\psi_s$) results were obtained, which are
consistent with the $\lambda$6\ cm observables
$PI_{{\mathrm{on}}}/PI_{{\mathrm{off}}}$ and $PA_{{\mathrm{on}}} -
PA_{{\mathrm{off}}}$. We then used these qualified ($f$, $c$,
$\psi_s$) to predict $\lambda$11\ cm $PI_{{\mathrm{on}}}$ ($PI =
\sqrt{U^2 + Q^2}$) and $PA_{{\mathrm{on}}}$ ($PA =
\frac{1}{2}~atan\frac{U}{Q}$) based on the following equation
\citep[for details see][]{Sun07}:

\begin{equation}
\centering
\displaystyle{
\left\{
\begin{array}{ll}
\displaystyle
U_{\mathrm{on}}=\mathit{f} PI_{\mathrm{bg}} {\rm sin}2\psi_s\ , \\ \displaystyle
\vspace{0.5mm}
Q_{\mathrm{on}}=PI_{\mathrm{fg}}+\mathit{f} PI_{bg}{\rm cos}2\psi_s.
&
\end{array}
\right.
}
\label{eq2}
\end{equation}

\noindent Here, $PI_{\rm fg,11cm}$ = $PI_{\rm off, 11cm} \times c$ and
$PI_{\rm bg, 11cm} = PI_{\rm off, 11cm} \times (1 -c)$, where
$\lambda$11\ cm $PI_{\rm off, 11cm}$ was calculated as $PI_{\rm off,
  11cm} = PI_{\rm off, 6cm}\ (\frac{\rm 2639~MHz}{\rm
  4800~MHz})^{\beta} = {\rm 65.8~mK~T}_{b}$, with a spectral index
$\beta = -3$ as obtained in Sect.~2.4. This agrees with the measured
values at the edges of the $\lambda$11\ cm $PI$ profile in
Fig.~\ref{profile}.  The rotation angles at $\lambda$11\ cm were
calculated as $\psi_{11cm} =
\psi_{6cm}~(\frac{0.114}{0.0625})^{2}$. We compared the predicted
values with the $\lambda$11\ cm observational results within 1$\sigma$
errors, which further filtered the ($f$, $c$, $\psi_s$) data to agree
with both $\lambda$6\ cm and $\lambda$11\ cm observations. From all
the remaining results, we found the parameter $c$ to be rather well
constrained within 0.8 to 0.9 (see Table~\ref{t2}) and largely
independent of $f$ variations. This is consistent with the dominant
foreground contribution towards the W4 superbubble. The reduction of
the polarized background emission is mathematically possible in
various ways. Here, $\psi_{s}$ is the most interesting parameter and
is also well constrained, which enables us to make a reasonable
estimate of the magnetic field and assess its errors. In Table~2, we
listed the average values of $f$, $c$, and $\psi_{s}$ for each column
in the Faraday screen, including standard deviation. We searched for
the best fit ($f$, $c$, $\psi_s$) for each column by the least
$\chi^2$ method. The difference between the average and the best fit
value ($f$, $c$, $\psi_s$) is small, with the maximum of
1.9$\sigma$. In the following, we used the averaged $\psi_s$ and its
standard deviation to derive the magnetic field within the western
shell of the W4 superbubble.
 
The $\lambda$21\ cm observations cannot be used for the model fitting
since depolarization occurs at $\lambda$21\ cm even by smaller RMs
within the shell. However, it can be examined to verify the fit
result. Considering the small-scale RM structures as shown by the
$\lambda$11\ cm data, total depolarization could be expected for the
background $PI$ at $\lambda$21\ cm passing through the bubble's
shell. In fact, the interferometric $\lambda$21\ cm image by
\citet{West07} shows the depolarization along the W4 superbubble
shells, which is almost masked in the single-dish polarization image
as discussed in Sect.~3.2. In the \citet{West07} interferometric map,
the depolarization averaged for a 10$\arcmin$ wide latitude stripe is
about 25~mK T$_{b}$ along the eastern and about 10~mK T$_{b}$ along
the western shell. The single-dish map polarization level of about
500~mK T$_{b}$ in the eastern and about 350~mK T$_{b}$ in the western
shell area imply an upper limit of about 0.97 for $c$ in the western
shell and about 0.95 in the eastern shell. The large difference in the
polarization emission levels in the interferometric and in the
single-dish maps means that a quantitative analysis of the
interferometric polarization data without large-scale correction
cannot be made to estimate the field strength. We concluded that the
$\lambda$21\ cm polarization data agree with the parameters of the
Faraday screen model along the western shell based on the
$\lambda$6\ cm and $\lambda$11\ cm observations.

For an estimate of $B_{//}$ within the western shell, the thermal
electron density n$_{e}$, its filling factor $f_{n_{e}}$, and the path
length $l$ within the bubble's shell are needed. Emission measure
($EM$), which is the integral of $f_{n_{e}}$ and n$_{e}$$^2$ over the
path length $dl$ can be calculated, because it is related to
optically-thin thermal radio continuum emission by the following
equation \citep{Wilson13}:

\begin{equation}
\centering 
\rm (\frac{\it T_{b}}{K}) = 8.235\times10^{-2}(\frac{\it T_{e}}{K})^{-0.35}(\frac{\nu}{GHz})^{-2.1}(\frac{\it EM}{pc~cm^{-6}})a(\nu,T),
\label{eq3}
\end{equation}

\noindent where, {\it T$_{b}$} is the radio continuum brightness
temperature measured at frequency $\nu$, a($\nu$,T) is a factor close
to 1. We assumed {\it T$_{e}$} = 8\,000~K for the ionized Galactic
interstellar medium in general. Based on the $\lambda$6\ cm
observation, we deduced the $EM$s within the western shell of the W4
superbubble. However, from Fig.~\ref{profile} we saw a large-scale
emission component of about 1.9~mK T$_{b}$ at $\lambda$6\ cm besides
thermal emission $I$ from the western shell. After subtracting this
large-scale component, $EM$s of the central eight columns within the
screen area were calculated based on Eq.~\ref{eq3}, as listed in Table
2.

To verify the $EM$ results that we obtained from the radio continuum
data, H$\alpha$ measurements were used to estimate the $EM$s
independently following \citet{Haffner98} and \citet{Finkbeiner03} by
using

\begin{equation}
\centering
EM =2.75~T_{4}^{0.9}~I_{H\alpha}~\exp[2.44E(B-V)].
\label{eq4}
\end{equation}

\noindent \citet{Dennison97} measured H$\alpha$ emission for several
positions towards the W4 superbubble. At their position K ($\ell \sim
133\fdg0$, $b \sim 5\fdg4$), they observed an intensity of
5.8~Rayleigh. \citet{Reynolds01} measured 6.9$\pm$0.5 Rayleigh at the
position $\ell = 132\fdg8, b = 5\fdg1$. Both directions are within the
western shell region of the W4 superbubble. The extinction parameter
$E(B-V),$ which must be taken into account, is in general difficult to
obtain and often ignored for regions outside the Galactic plane. Two
stars, Hilt~266 ($\ell = 132\fdg2, b =3\fdg1$) and Hilt~338 ($\ell =
134\fdg2, b =3\fdg0$), were found near the W4 superbubble with
distances of 2.5~kpc and 2.3~kpc, respectively. They have a measured
extinction value of $E(B-V)$ = 0.66 \citep{Hiltner56}. Based on
Eq.~\ref{eq4} and assuming T = 8\,000~K, we derived the $EM$s for the
two positions as 65~$\rm pc\ cm^{-6}$ and 77.7$\pm$5.6~$\rm
pc\ cm^{-6}$, which are consistent with those based on radio continuum
data in Table~2.

The path lengths for the central eight columns (9th -16th) across the
western shell of the W4 superbubble are related to the geometry, where
we assume a uniform shell. The largest RM corresponds to the longest
path length \citep{Vallee82}, which is corresponding to the 13th
column in our case (see Fig.~\ref{profile} and Table~2), and defines
the inner radius of the shell. We also considered the case of a
thermal shell, where the strongest total intensity emission is at the
12th column, and included the results in Table~2. The angle between
the central axis of the W4 superbubble and the 13th column is about
1$\fdg$7, and the angle between the 13th column and the outer edge of
the western shell is 10$\farcm$5. Then for a distance of 2.2~kpc, we
calculated the inner and outer radii of the shell as 65~pc and 72~pc,
respectively. With a column size of 3$\arcmin$, the path lengths of
the central eight columns were deduced. The electron densities and the
strength of the line-of-sight component of the magnetic field $B_{//}$
can be obtained following the equations given by
\citet{Harvey-Smith11}, where the filling factor $f_{n_{e}}$ is the
only unknown parameter:

\begin{equation}
\centering
n_{e} = \sqrt{\frac{EM}{\mathit{f_{n_{e}}}l}}~ {\rm cm^{-3}},
\label{eq5}
\end{equation}

\begin{equation}
\centering
B_{//} = \frac{RM}{0.81\sqrt{EM}\sqrt{\mathit{f_{n_{e}}}l}}~ {\rm \mu G}.
\label{eq6}
\end{equation}

\begin{figure*}
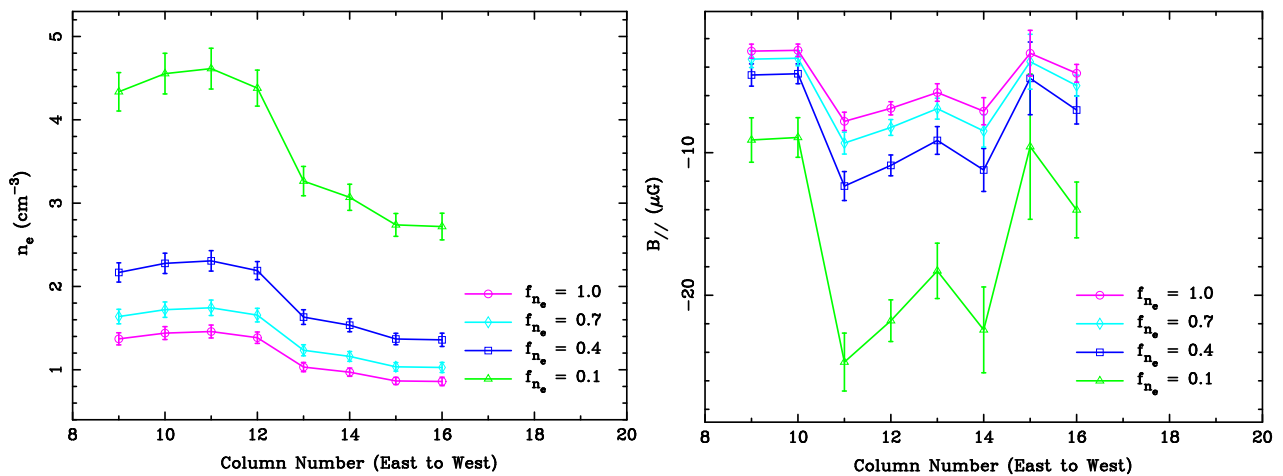
 
\centering
\includegraphics[angle=-90, width=0.45\textwidth]{24952fg7a.ps}
\includegraphics[angle=-90, width=0.45\textwidth]{24952fg7b.ps}
\caption{{\it Left panel}: electron density n$_{e}$ across the western
  shell of the W4 superbubble for filling factors $f_{n_e}$ = 0.1
  (green triangle), 0.4 (blue square), 0.7 (cyan diamond), and 1.0
  (magenta circle). {\it Right panel}: the strength of the
  line-of-sight component of the B-field across the western shell for
  different filling factors $f_{n_e}$.}
\label{fneb}
\end{figure*}

The filling factor is often assumed to be 1.0, as in \citet{West07},
while \citet{Harvey-Smith11} used $f_{n_{e}} = 0.1$ for the sample of
\ion{H}{II} regions they studied.  Since no well-determined value has
been reported for superbubbles so far, we calculated n$_{e}$ and
$B_{//}$ based on Eqs.~\ref{eq5} and~\ref{eq6}, for $f_{n_{e}}$ = 0.1,
0.4, 0.7, and 1.0. We used the standard error propagation to estimate
the errors in n$_{e}$ and $B_{//}$. The uncertainty of n$_{e}$ results
from the error of $EM$, which stems from the $\lambda$6\ cm total
intensity measurements. \citet{Sun07} quote the maximum uncertainty
of 10\% for the $\lambda$6\ cm survey data.  Besides the uncertainty
of $EM$, the errors in $B_{//}$ also depend on the uncertainties in
RM. The standard deviations used for this purpose were listed in
Table~2.

We showed the n$_{e}$ and $B_{//}$ profiles across the western shell
of the W4 superbubble in Fig.~\ref{fneb}. Owing to the
$1/\sqrt{f_{n_e}}$ dependence, $f_{n_e}$ = 0.1 results in high
electron density and a large B field, while the values for $f_{n_e}$ =
0.4, 0.7, and 1.0 are much smaller. To distinguish which case is more
realistic, we made a simple estimate according to the magnetic flux
conservation.  For the W4 superbubble, the swept-up magnetic field
should be a factor of about 5 higher than the undisturbed Galactic B
field when we consider a ring with an inner radius of 65~pc and an
outer radius of 72~pc. Using RM data from pulsars, \citet{Han06} find
that the strength of the large-scale regular radial B field of the
Galaxy decreases from the inner Galaxy (4~$\mu$G) to the solar
neighbourhood (2~$\mu$G). If a turbulent field of the same order is
considered, the field strength in the W4 superbubble area is about
3~$\mu$G. The 3-D simulation of \citet{Sun08} and \citet{Sun09} where
both regular and turbulent B fields were included gave a similar value
of B$_{tot}$ of around 2.4~$\mu$G. Thus the enhanced line-of-sight
B-field strength in the western shell should not exceed 15~$\mu$G by
much, otherwise an amplification mechanism is needed to explain a
larger B field for $f_{n_e}$ = 0.1. If a B-field amplification can be
excluded, the filling factor $f_{n_e}$ in the western shell of the W4
superbubble is likely within the range of 0.4 to 1.0. We limit the
following discussion to the case that the filling factor $f_{n_e}$ is
0.4, 0.7, and 1.0, where $f_{n_e}$ = 1.0 gives the lower limit of the
electron density n$_{e}$ and the strength of the line-of-sight
component of the magnetic field B$_{//}$.  When considering that the
line-of-sight passes solely through the western shell (13th to 16th
column), we obtained the average B$_{//}$ to be
$-$5.0($\pm$10\%)~$\mu$G ($f_{n_e}$ = 1.0) to $-$8.0($\pm$10\%)~$\mu$G
($f_{n_e}$ = 0.4), and the electron density n$_{e}$ is between
1.0($\pm$5\%)~cm$^{-3}$ ($f_{n_e}$ = 1.0) and 1.5($\pm$5\%)~cm$^{-3}$
($f_{n_e}$ = 0.4).

\citet{West07} used a different analysis method to derive the
line-of-sight component of the magnetic field strength in the shell of
the W4 supperbubble. Based on the observed depolarization amount, they
estimated the PA difference from adjacent lines passing through the
shell.  From the Faraday screen model, we found the RMs within the
shell for the 13th to 16th column between $-$70~rad\ m$^{-2}$ and
$-$300~rad\ m$^{-2}$, which causes one to four full rotations of PAs
at $\lambda$21\ cm.  \citet{West07} quote the $\lambda$21\ cm
differential angle of 60$\degr$ and calculated a line-of-sight B-field
strength between 3.4~$\mu$G and 9.1~$\mu$G from the different path
lengths through the shell. The polarization intensity, polarization
angles, and their differences are, however, influenced by the
large-scale polarized emission, which is incompletely observed by the
DRAO interferometer. \citet{Sun07} quote the observed Stokes parameter
$U$, $Q$ from a passive Faraday screen as $U = PI_{fg} {\rm
  sin}2\psi_{fg} + \mathit{f}PI_{bg}{\rm sin}(2\psi_{bg}+2\psi_{s})$,
$Q = PI_{fg}{\rm cos}2\psi_{fg} + \mathit{f}PI_{bg}{\rm
  cos}(2\psi_{bg}+2\psi_{s})$, where $fg$ and $bg$ denote foreground
and background components of $PI$ and $PA$. Here, $f$ is the
depolarization factor and $\psi_{s}$ is the Faraday rotation angle
within the screen. For the W4 superbubble, the foreground and
background $PA$ angles are close to zero (see Sect.~4.1.), then $U$
and $Q$ were calculated as $U = \mathit{f}PI_{bg}{\rm sin}2\psi_{s}$
and $Q = PI_{fg} + \mathit{f}PI_{bg}{\rm cos}2\psi_{s}$
(Eq.~2). Therefore, the observed depolarization amount in $PI$ and the
$PA$ differences for neighbouring lines through the shell depend not
only on the difference in $\psi_{s}$, which is needed for a B-field
calculation, but also on $PI_{fg}$ and $PI_{bg}$. Without taking this
into account, the B-field result depends in some way on the
large-scale polarization fraction included in the data. Only if the
Faraday screen is very local and $PI_{fg}$ = 0~mK T$_{b}$ does this
method give the correct results. However, this is not the case for the
W4 superbubble.

\subsection{The north-eastern extension}

\begin{figure*} 
\centering
\includegraphics[angle=-90, width=0.45\textwidth]{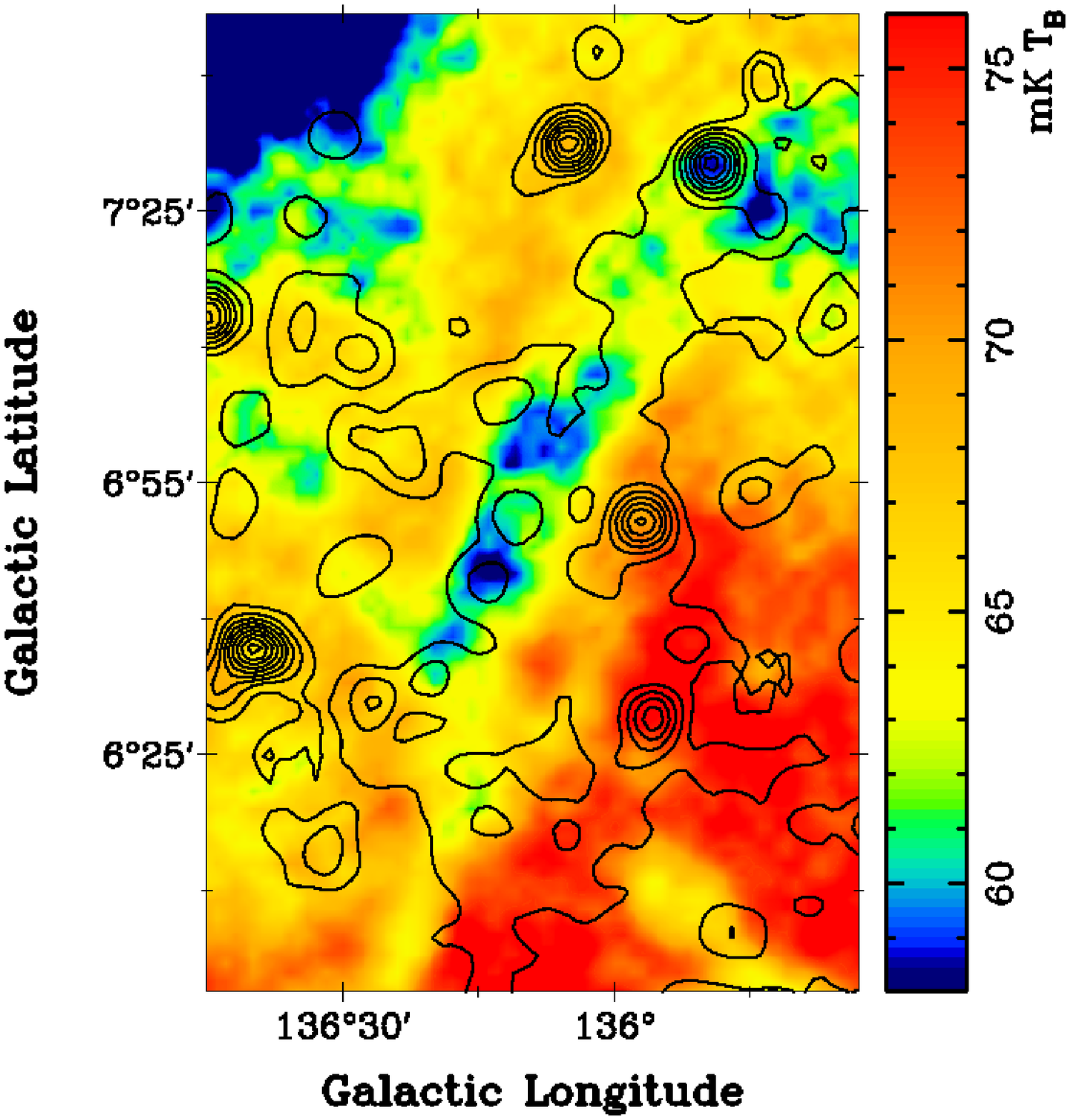}
\includegraphics[angle=-90, width=0.45\textwidth]{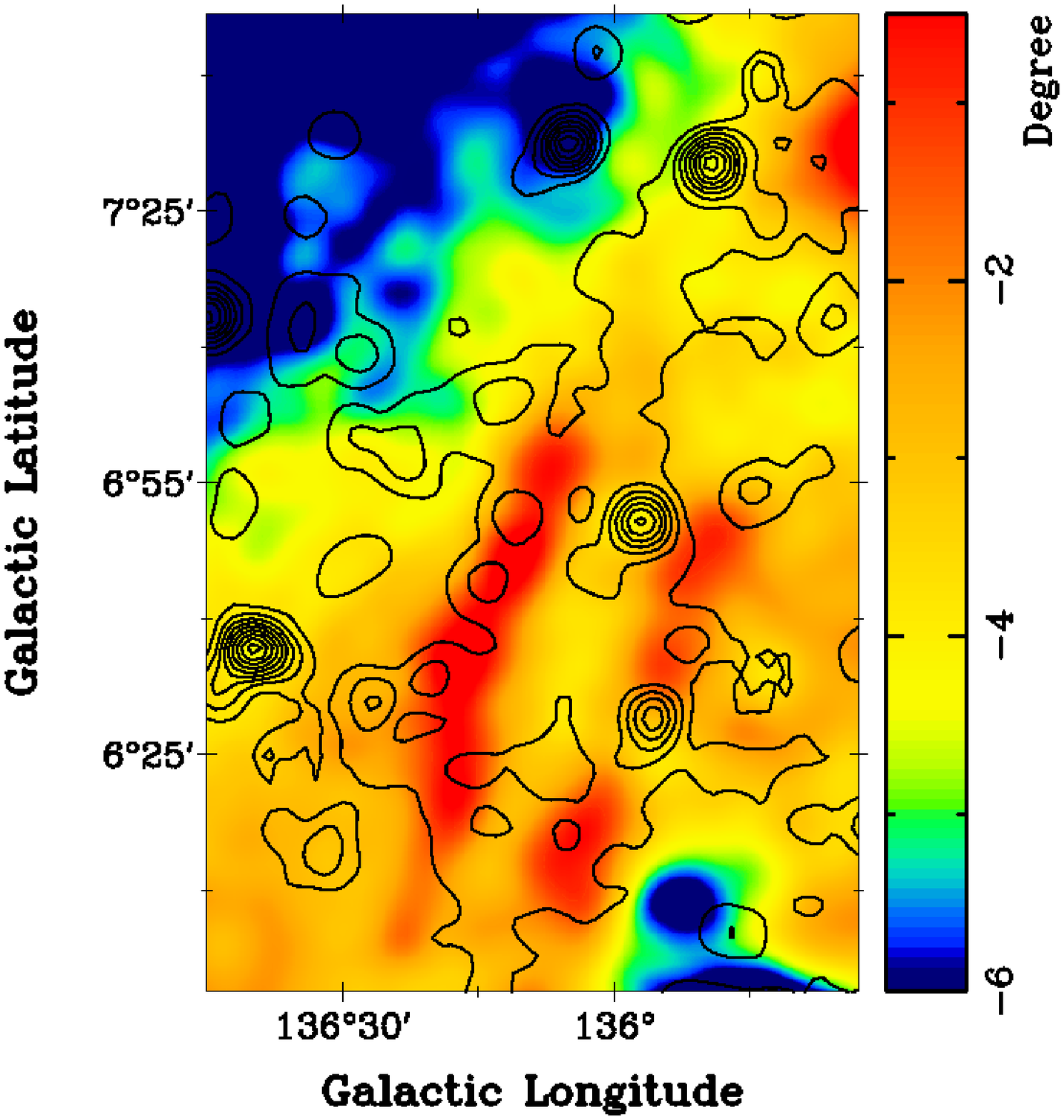}
\caption{$\lambda$11\ cm $PI$ ({\it left panel}) and $PA$ ({\it right
    panel}) of the north-eastern shell extension overlaid by total
  intensity contours. The contour levels and the angular resolution of
  the images are the same as in Fig.~\ref{ipi}.}
\label{neshell}
\end{figure*}

For better understanding of the magnetic field configuration and the
strength of the W4 superbubble, we searched for more parts of shells
and arcs, where Faraday effects could be studied in terms of the
described Faraday screen model. We note that it is difficult to find
unconfused areas with good quality ``on'' and ``off'' data. However, a
section of the north-eastern extension located above the major eastern
shell is noted in the $\lambda$11\ cm polarization image (see
Fig.~\ref{neshell}), which we found to be suitable for modelling. The
$\lambda$6\ cm data show no evidence of depolarization in this region,
indicating that the RM is smaller than in the western part of the W4
superbubble.  We again applied the Faraday screen model and obtained
$f \sim 0.87$, $c \sim 0.86$ and a positive RM of about 55~rad
m$^{-2}$ in this area. The fitted parameters predict $PI_{\rm
  on}/PI_{\rm off} = 0.97$ and $PA_{\rm on} - PA_{\rm off} = 1\fdg5$
for the $\lambda$6\ cm, and $PI_{\rm on}/PI_{\rm off} = 0.89$ and
$PA_{\rm on} - PA_{\rm off} = -3\fdg9$ for the $\lambda$21\ cm
observations, which generally agrees with the non-detection of
noticeable Faraday rotated structures in both maps. Following the
procedures introduced above, we obtained a lower electron density of
n$_{e}$ = 0.5 ($f_{n_e}$ = 1.0) - 0.9~cm$^{-3}$ ($f_{n_e}$ = 0.4) with
1$\sigma$ error of 15\%, and a weaker line-of-sight component of the
magnetic field B$_{//}$ = 3.1 ($f_{n_e}$ = 1.0) - 4.9~$\mu$G
($f_{n_e}$ = 0.4) with 1$\sigma$ error of 30\% within the upper
north-eastern shell extension.

\section{Discussion and summary}

We have shown that strong regular magnetic fields exist in the W4
superbubble's shells. A study of the process of creating this magnetic
field configuration is beyond the scope of this paper, while numerical
simulations \citep[e.g.][]{Tomisaka98, Avillez05} provide hints of an
evolutionary scenario. To explain the observations, it seems most
likely that the originally plane-parallel magnetic fields were lifted
up and compressed during the expansion of the W4 superbubble.

The calculation of the total magnetic field strength in the W4
superbubble shell depends on its orientation relative to the line of
sight.  From the discovery of atomic hydrogen associated to the W4
superbubble, \citet{Normandeau96} concluded that the W4 superbubble is
inclined towards the observer. \citet{Lagrois09} studied this
inclination in detail and found that the inclination angle is between
9$\degr$ and 27$\degr$. In addition, we measured an inclination angle
of 18$\degr$ with respect to the Galactic north pole for the western
shell (see Fig.~\ref{ipi}). If the magnetic field in the W4
superbubble is directed along its shell and if taking its geometry
into account, the deprojected regular {\it B$_{tot}$} field in the
western shell can be deduced as $B_{tot} =
B_{//}~\sqrt{1+(\frac{1}{{\rm tan}\theta\ {\rm cos}18\degr})^2} = (2.3
- 6.7)~B_{//}$, where $\theta$ is the inclination angle between
9$\degr$ and 27$\degr$.

\begin{table*}[thp]
\caption{Thermal pressure versus magnetic pressure in the western
  shell of the W4 superbubble for different filling factors $f_{n_e}$
  and the inclination angles of $\theta = 27\degr$, $\theta =
  18\degr$, and $\theta = 9\degr$.}
\label{tab3}
\vspace{-1mm}
\centering
\begin{tabular}{rcccc}
\hline\hline
\multicolumn{1}{c}{Column No.}    &\multicolumn{1}{c}{P$_{ther}$}  & \multicolumn{1}{c}{P$_{mag}$, $\theta = 27\degr$}  & \multicolumn{1}{c}{P$_{mag}$, $\theta = 18\degr$}  &\multicolumn{1}{c}{P$_{mag}$, $\theta = 9\degr$} \\
                                 & (10$^{-12}$\ dyn\ cm$^{-2}$)   & (10$^{-12}$\ dyn\ cm$^{-2}$)            & (10$^{-12}$\ dyn\ cm$^{-2}$)            & (10$^{-12}$\ dyn\ cm$^{-2}$) \\
\hline                          
 $f_{n_e} = 0.4$           &                               &                      &                      & \\
13th                       &3.6$\pm$0.2                   &17.5$\pm$3.7          &38.2$\pm$8.1          &150.0$\pm$31.9   \\
14th                       &3.4$\pm$0.2                   &26.3$\pm$7.1          &57.4$\pm$15.4         &225.5$\pm$60.5   \\
16th                       &3.0$\pm$0.2                   &10.3$\pm$2.9          &22.4$\pm$6.3          &88.1$\pm$24.6   \\
\hline
 $f_{n_e} = 0.7$           &                               &                      &                      & \\
13th                       &2.7$\pm$0.1                    &10.0$\pm$2.1         &21.8$\pm$4.6          &85.7$\pm$18.2   \\
14th                       &2.6$\pm$0.1                    &15.0$\pm$4.0         &32.8$\pm$8.8          &128.9$\pm$34.6   \\
16th                       &2.3$\pm$0.1                    &5.9$\pm$1.6          &12.8$\pm$3.6          &50.3$\pm$14.1   \\
\hline
 $f_{n_e} = 1.0$           &                               &                      &                      & \\
13th                       &2.3$\pm$0.1                    &7.0$\pm$1.5          &15.3$\pm$3.2          &60.0$\pm$12.7   \\
14th                       &2.1$\pm$0.1                    &10.5$\pm$2.8         &23.0$\pm$6.2          &90.2$\pm$24.2   \\
16th                       &1.9$\pm$0.1                    &4.1$\pm$1.1          &9.0$\pm$2.5           &35.2$\pm$9.8   \\
\hline
\vspace{1mm}
\end{tabular}\\
\end{table*}

Thermal and magnetic pressure, besides gravity, ram pressure, and
turbulence are key input parameters for superbubble evolution
simulations.  Thermal pressure is defined as $P_{ther} =
2n_{0}kT_{e}$, where $\rm n_{0} = n_{e}$ for total ionization, k is
the Boltzmann constant, and $T_{e}$ is 8\,000~K as assumed above.
Magnetic pressure is calculated as $P_{mag} = B_{tot}^{2}/8\pi$. For
each column within the western shell of the W4 superbubble shell
(13th, 14th, and 16th columns, the 15th column is excluded because of
the poor constraints on the B-field), we calculated the thermal and
magnetic pressure based on the results shown in Fig.~\ref{fneb} and
made a comparison in Table~3. For the inclination angle of $\theta$ =
27$\degr$, the magnetic pressure $P_{mag}$ is generally several times
of the thermal pressure $P_{ther}$, and $P_{mag}$ becomes overwhelming
when the inclination angle $\theta$ approaches to 9$\degr$.

For the north-eastern extension in the higher latitude region,
$P_{mag}$ and $P_{ther}$ are comparable when $\theta \sim 27\degr$,
regardless of $f_{n_e}$ = 0.4 or 1.0. $P_{mag}$ dominates $P_{ther}$
if $\theta$ approaches 9$\degr$, even for $f_{n_e}$ = 1.0. A more
precise inclination angle would help to better constrain the physical
properties of the W4 superbubble.

We found a positive RM on the eastern side and a negative RM on the
western side of the W4 superbubble. This is expected for the scenario
described above, where the W4 superbubble expands and breaks out of
the Galactic plane and lifts up the magnetic field, which runs
clockwise (viewed from the north Galactic pole) in the Perseus arm
\citep{Han06}. Then, for the eastern shell, the field lines will go up
and for the western shell downwards. Because the W4 superbubble tilts
towards us \citep{Normandeau96, Lagrois09}, the line-of-sight
component of the magnetic field points away in the western shell of
the W4 superbubble, resulting in a negative RM, and towards us in the
eastern shell, where a positive RM is observed.

Radio continuum and polarization observations at $\lambda$6\ cm,
$\lambda$11\ cm, and $\lambda$21\ cm have been made to study the radio
emission properties and to estimate the magnetic fields of the W4
superbubble.  With the flat radio continuum spectrum found between
$\lambda$6\ cm, $\lambda$11\ cm, and $\lambda$21\ cm, we confirm the
thermal origin of the radio continuum emission of the W4
superbubble. Polarized emission shows dramatically morphological
differences from wavelength to wavelength. With the advantage of being
less affected by Faraday rotation, the $\lambda$6\ cm and
$\lambda$11\ cm polarization data were used for the estimates of the
line-of-sight component of the magnetic field within the western shell
of the W4 superbubble by a passive Faraday screen model. Considering
the thermal electron filling factor $f_{n_e}$, the radio continuum
observations and a simple geometric assumption result in an electron
density of n$_{e}$ $\sim$ 1.0 $\times$ $1/\sqrt{f_{n_e}}$~cm$^{-3}$,
and the line-of-sight component of the magnetic field $B_{//} \sim$
$-$5.0 $\times$ $1/\sqrt{f_{n_e}}$~$\mu$G (pointing away from us) for
the western shell of the W4 superbubble, where the typical error is
about 5\% for n$_{e}$ and 10\% for B$_{//}$. Based on a simple
estimate, we found that $f_{n_e}$ likely has a value greater than 0.1
assumed for \ion{H}{II} regions \citep{Harvey-Smith11}.  Being related
to the inclination angle of the superbubble with respect to the plane
of the sky, a total magnetic field is found above 12~$\mu$G. This
results in a magnetic pressure, which is one or two orders of
magnitude higher than the thermal pressure in the western shell of the
W4 superbubble. The $\lambda$11\ cm polarization data allow a model
fit of the weaker RM in the high-latitude north-eastern shell, where
we find a lower electron density of n$_{e}$ = 0.5 to 0.9\ cm$^{-3}$
and $B_{//}$ of 3.1 $-$ 4.9~$\mu$G for filling factors $f_{n_e}$ of
1.0 or 0.4, respectively. This means that the magnetic and thermal
pressure might be comparable in the upper parts of the
superbubble. The RM sign reverses as expected for a scenario where the
Galactic magnetic field is pushed out of the plane by the expanding W4
superbubble, which is tilted towards us. These results are expected to
constrain magneto-hydrodynamical simulations of the W4 superbubble and
superbubbles in general.

\begin{acknowledgements}
XYG and JLH are supported by the National Natural Science foundation
of China (11303035, 11473034) and the Partner group of the MPIfR at
NAOC in the framework of the exchange programme between MPG and CAS
for many bilateral visits. XYG acknowledges financial support by the
MPG, by Michael Kramer during his stay at the MPIfR, Bonn, and the
Young Researcher Grant of National Astronomical Observatories, Chinese
Academy of Sciences. This research is based in part on observations
with the Effelsberg 100-m telescope of the MPIfR. We would like to
thank the anonymous referee for helpful comments and suggestions.
\end{acknowledgements}

\bibliographystyle{aa}
\bibliography{bbfile}

\end{document}